\documentclass{99-Styles/CCAP}
\usepackage{afterpage}
\usepackage{longtable}
\usepackage{rotating}
\usepackage{textcomp}
\usepackage{lineno}
\usepackage{tikz}
\usepackage{times}
\usepackage{bm}         
\usepackage{amsmath,amssymb,bm,slashed} 
\usepackage{amssymb}    
\usepackage{graphicx}   
\usepackage{verbatim}   
\usepackage{color}      
\definecolor{RedViolet}{cmyk}{0.60, 0.99, 0.99, 0.0}
\definecolor{BlueViolet}{cmyk}{0.90, 0.90, 0.0, 0.0}
\definecolor{DarkBlue}{cmyk}{0.90, 0.65, 0.0, 0.0}
\definecolor{DarkGreen}{cmyk}{0.95, 0.25, 0.95, 0.25}
\definecolor{DarkYellow}{cmyk}{0.0, 0.25, 0.95, 0.25}
\usepackage{subfigure}  
\usepackage{hyperref}   
\usepackage{longtable}
\usepackage{enumerate}

\usepackage{fmtcount}   

\usepackage[square,comma,numbers,sort&compress]{natbib}

\usepackage[detect-all]{siunitx}  
\usepackage{enumitem}
\usepackage{blindtext}
\usepackage{wrapfig}
\usepackage{tabularx}

\newcommand{\parenbar}[1]{\ensuremath{\overset{\scriptscriptstyle{(-)}}{#1}}}

\newcommand{\lsim}      {\mbox{\raisebox{-0.4ex}{$\;\stackrel{<}{\scriptstyle \sim}\;$}}}

\usepackage{bibentry}
\nobibliography*

\begin{document}

\makeatletter

\newcommand{\bra}[1]{\ensuremath{\langle #1 |}}   
\newcommand{\ket}[1]{\ensuremath{| #1 \rangle}}   
\newcommand{\bigbra}[1]{\ensuremath{\big\langle #1 \big|}}   
\newcommand{\bigket}[1]{\ensuremath{\big| #1 \big\rangle}}   
\newcommand{\amp}[3]{\ensuremath{\left\langle #1 \,\left|\, #2%
                     \,\right|\, #3 \right\rangle}}  
\newcommand{\sprod}[2]{\ensuremath{\left\langle #1 |%
                     #2 \right\rangle}}  
\newcommand{\ev}[1]{\ensuremath{\left\langle #1 %
                     \right\rangle}} 
\newcommand{\ds}[1]{\ensuremath{\! \frac{d^3#1}{(2\pi)^3 %
                     \sqrt{2 E_\vec{#1}}} \,}} 
\newcommand{\dst}[1]{\ensuremath{\! %
                     \frac{d^4#1}{(2\pi)^4} \,}} 
\newcommand{\tr}{\text{tr}}
\newcommand{\sgn}{\text{sgn}}
\newcommand{\diag}{\text{diag}}
\newcommand{\BR}{\text{BR}}

\renewcommand{\vec}[1]{{\mathbf{#1}}}
\renewcommand{\Re}{{\text{Re}}}
\renewcommand{\Im}{{\text{Im}}}
\newcommand{\iso}[2]{{\ensuremath{{}^{#2}}\ensuremath{\rm #1}}}
\newcommand{\eps}{{\ensuremath{\epsilon}}}
\newcommand{\draftnote}[1]{{\bf\color{red} \MakeUppercase{#1}}}
\newcommand{\panm}[1]{{\color{blue} #1}}
\providecommand{\abs}[1]{\lvert#1\rvert}
\providecommand{\norm}[1]{\lVert#1\rVert}

\def\parenbar{\mathpalette\p@renb@r}
\def\p@renb@r#1#2{\vbox{%
  \ifx#1\scriptscriptstyle \dimen@.7em\dimen@ii.2em\else
  \ifx#1\scriptstyle \dimen@.8em\dimen@ii.25em\else
  \dimen@1em\dimen@ii.4em\fi\fi \offinterlineskip
  \ialign{\hfill##\hfill\cr
    \vbox{\hrule width\dimen@ii}\cr
    \noalign{\vskip-.3ex}%
    \hbox to\dimen@{$\mathchar300\hfil\mathchar301$}\cr
    \noalign{\vskip-.3ex}%
    $#1#2$\cr}}}

%
\providecommand{\anmne}{\mbox{$\bar\nu_{\mu} \rightarrow \bar\nu_e$}} 
\providecommand{\nmne}{\mbox{$\nu_{\mu}\rightarrow\nu_e$}} 
\providecommand{\anm}{\mbox{$\bar\nu_\mu$}} 
\providecommand{\nm}{\mbox{$\nu_\mu$}}
\providecommand{\nue}{\mbox{$\nu_e$}} 
\providecommand{\ane}{\mbox{$\bar\nu_e$}} 
\providecommand{\enu}{\mbox{$E_\nu$}}
\providecommand{\piz}{\mbox{$\pi^0 $}}
\providecommand{\pip}{\mbox{$\pi^+$}} 
\providecommand{\pim}{\mbox{$\pi^-$}}

\parindent 10pt
\pagenumbering{roman}
\setcounter{page}{1}
\pagestyle{plain}

\thispagestyle{empty}

\leftline{Final} \vspace{-0.4cm} \rightline{\today} \vspace{0.5cm}

\noindent{\bf\LARGE 
  Neutrinos from Stored Muons (nuSTORM) \\
}
\noindent\textit{Submitted to the Snowmass 2021 DPF Community Planning Exercise}


\vspace{1.0cm}
\noindent{
  L.~~Alvarez~Ruso${}^{1}$,
  T.~Alves${}^{2}$,
  S.~Boyd${}^{3}$,
  A.~Bross${}^{4}$,
  P.R.~Hobson${}^{5}$,
  P.~Kyberd${}^{6}$,
  J.B.~Lagrange${}^{7}$, 
  K.~Long${}^{2,7}$,
  X.-G.~Lu${}^{3}$,
  J.~Pasternak${}^{2,7}$,
  M.~Pfaff${}^{2,8}$,
  C.~Rogers${}^{7}$ 
  \renewcommand{\thefootnote}{\fnsymbol{footnote}}
  on behalf of the nuSTORM collaboration\footnote[2]{
    The nuSTORM collaboration is presented in the addendum.
  }

}

\begin{flushleft}

  \begin{minipage}{15cm}
    {\em\footnotesize
      ${}^{1}$ Instituto de Fisica Corpuscular (IFIC), Centro Mixto
        CSIC-UVEG, Edificio Institutos Investigación, Paterna,
        Apartado 22085, 46071 Valencia, Spain \\
      ${}^{2}$ Physics Department, Blackett Laboratory, Imperial
        College London, Exhibition Road, London, SW7 2AZ, UK \\
      ${}^{3}$ Department of Physics, University of Warwick, Coventry,
        CV4 7AL, UK \\
      ${}^{4}$ Fermilab, P.O. Box 500, Batavia, IL 60510-5011, USA \\
      ${}^{5}$ Department of Physics and Astronomy, Queen Mary University of London,
        London, E1 4NS, UK \\
      ${}^{6}$ College of Engineering, Design and Physical Sciences,
        Brunel University, Uxbridge, Middlesex, UB8 3PH, UK \\
      ${}^{7}$ STFC, Rutherford Appleton Laboratory, Harwell Campus,
       Didcot, OX11 0QX  \\
      ${}^{8}$ Department of Physics, Technical University Munich (TUM), James-Franck-Str. 1,
       D-85748 Garching, Germany \\
    }
  \end{minipage}

  \vspace{5mm}
  \noindent{\small }
  
\end{flushleft}

\section*{\color{RedViolet}\Large Executive summary}

The 2020 Update of the European Strategy for Particle Physics
(ESPP)~\cite{EuropeanStrategyGroup:2020pow} recommended that muon beam
R\&D should be considered a high-priority future initiative and that a
programme of experimentation be developed to determine the neutrino
cross-sections required to extract the most physics from the DUNE and
Hyper-K long-baseline experiments.
The ENUBET~\cite{ENUBET:WWW,ENUBET:EUWWW,Torti:2020yzn} and
nuSTORM~\cite{nuSTORM:WWW,Ahdida:2020whw} 
collaborations have begun to work within and alongside the CERN
Physics Beyond Colliders study group~\cite{PBC:WWW} and the
international Muon Collider collaboration~\cite{iMC:WWW} to carry out
a joint, five-year R\&D programme to deliver a
detailed plan for the implementation of an infrastructure in
which:
\begin{itemize}
  \item ENUBET and nuSTORM deliver the neutrino cross-section
    measurement programme identified in the ESPP and allow sensitive
    searches for physics beyond the Standard Model to be carried out;
    and in which
  \item A 6D muon ionisation cooling experiment is delivered as part
    of the technology development programme defined by the
    international Muon Collider collaboration.
\end{itemize}
This document summarises the status of development of the nuSTORM and
6D cooling experiments and identifies opportunities for collaboration
in the development of the initiative outlined above.

\subsection*{\textit{Elements of the proposed initiative:}}

\begin{description}
  \item \textbf{ENUBET} \\
    The ENUBET (Enhanced NeUtrino BEams from kaon Tagging; NP06)
    collaboration proposes a dedicated facility to measure
    $\nu_\mu$ and $\nu_e$ cross-sections precisely using a combination of
    monitored, narrow-band neutrino beams at the GeV energy scale and by
    instrumenting the meson-decay tunnel with a segmented calorimeter.
    The ENUBET approach is based on monitoring the production of
    large-angle positrons from $K^+ \rightarrow \pi^0 e^+ \nu_e$ (Ke3)
    decays in the decay tunnel.
    In addition, ENUBET will monitor muons produced in kaon and pion
    decays, thus providing a precise measurement of the $\nu_\mu$ flux.
    Due to the optimisation of the focusing-and-transport system of the 
    momentum-selected narrow-band beam of the parent mesons, the Ke3
    decay represents the main source of electron neutrinos.
    Furthermore, the positron rate may be used to measure the $\nu_e$ flux
    directly. 
    Consequently, the monitored $\nu_e$ beam will lower the uncertainties
    on the neutrino flux and flavour for a conventional beam from the
    current level of O(7\%-10\%) to $\sim 1\%$.
    Similar precision is expected for the $\nu_\mu$ flux, with the bonus
    that the neutrino energy will be determined with a precision
    of $\sim 10$\% at the single neutrino level by the “narrow-band off-axis
    technique”, i.e. using only the position of the $\nu_\mu$ interaction
    vertex. \\ 

  \item\textbf{nuSTORM} \\
    The Neutrinos from Stored Muons, nuSTORM, facility has been designed
    to deliver a definitive neutrino-nucleus scattering programme using
    beams of $\parenbar{\nu}\!\!_e$ and $\parenbar{\nu}\!\!_\mu$ from the decay
    of muons confined within a storage ring.
    The facility is unique, it will be capable of storing $\mu^\pm$
    beams with momentum of between 1\,GeV/c and 6\,GeV/c and a momentum
    spread of $\pm 16$\%.
    The neutrino beams generated will span neutrino energies from approximately
    300\,MeV to 5.5\,GeV.
    This will allow neutrino-scattering measurements to be made
    over the kinematic range of interest to the DUNE and Hyper-K collaborations.
    At nuSTORM, the flavour composition of the beam and the
    neutrino-energy spectrum are both precisely known.
    The storage-ring instrumentation will allow the neutrino flux to be
    determined to a precision of 1\% or better.
    By exploiting sophisticated neutrino-detector techniques such as
    those being developed for the near detectors of DUNE and Hyper-K,
    the nuSTORM facility will:
    \begin{itemize}
      \item Serve the future long- and short-baseline neutrino-oscillation
        programmes by providing definitive measurements of
        $\parenbar{\nu}\!\!_e A$ and $\parenbar{\nu}\!\!_\mu A$
        scattering cross-sections with percent-level precision;
      \item Provide a probe that is 100\% polarised and sensitive to
        isospin to allow incisive studies of nuclear dynamics and
        collective effects in nuclei;
      \item Deliver the capability to extend the search for light
        sterile neutrinos beyond the sensitivities that will be provided
        by the FNAL Short Baseline Neutrino (SBN) programme; and
      \item Create an essential test facility for the development of 
        muon accelerators to serve as the basis of a multi-TeV 
        lepton-antilepton collider and a Neutrino Factory.
      \end{itemize}
    To maximise its impact, nuSTORM should be implemented such that
    data-taking begins by $\approx 2030$ when the DUNE and Hyper-K
    collaborations will each be accumulating data sets capable of
    determining oscillation probabilities with percent-level
    precision. \\

  \item\textbf{Muon Collider demonstrator} \\
    Muon beams of high brightness have been proposed as the source of
    neutrinos at a Neutrino Factory and as the means to deliver multi-TeV
    lepton-antilepton collisions at a Muon Collider.
    In most of these proposals the muon beam is derived from pion decay as
    is proposed here for nuSTORM.
    nuSTORM, which will have beams with the highest ever stored-muon
    beam power, will allow many of the challenges associated with the
    muon storage ring in such facilities to be addressed, including:
    \begin{itemize}
      \item The complete implementation of a muon storage ring of large
        acceptance including the injection and extraction sections; and
      \item The design and implementation of instrumentation by which to
        determine the muon-beam energy and flux to 1\% or better.
        A novel polarimeter system will be required in order to determine
        the stored-muon energy and the energy spread.
    \end{itemize}
    
    The opportunity nuSTORM provides for the study of ionisation cooling
    is particularly important.
    The Muon Ionisation Cooling Experiment (MICE)~\cite{MICE:WWW} has
    demonstrated ionisation cooling in the 4-dimensional transverse
    phase space~\cite{MICE:Nature}.
    To prove the feasibility of a Muon Collider therefore requires a
    follow-on demonstration of ionisation cooling in the full
    six-dimensional (6D) phase space.
    The facility we propose to develop will be capable of delivering the
    required demonstration of 6D ionisation cooling.

\end{description}

\subsection*{\textit{Opportunity:}}

With their existing proton-beam infrastructure, CERN and Fermilab are
both uniquely well-placed to implement ENUBET, nuSTORM, and the
6D-cooling experiment as part of the required Muon Collider
demonstrator. 
The design of ENUBET, carried out within the framework of a European
Research Council funded design study, includes the precise layout of
the kaon/pion focusing beamline, photon veto and timing system as well
as the development and test of a positron tagger together with the
required electronics and readout.
The feasibility of implementing nuSTORM at CERN has been studied by
the CERN Physics Beyond Colliders study group while a proposal to site
nuSTORM at FNAL was developed for the last Snowmass study in 2013.
The FNAL study focused on the optimisation of the muon storage ring to
provide exquisite sensitivity in the search for sterile neutrinos.
In the Physics Beyond Colliders study, the muon storage ring was
optimised to carry out a definitive neutrino-nucleus scattering
programme using stored muon beams with momentum in the range 1\,GeV/c to
6\,GeV/c while maintaining its sensitivity to physics beyond the
Standard Model.

The study of nuSTORM is now being taken forward in the context of the
demonstrator facility required by the international Muon Collider
collaboration that includes the 6D muon ionisation cooling experiment.
The muon-beam development activity is being carried out in close
partnership with the ENUBET collaboration and the Physics Beyond 
Colliders Study Group.
In consequence we now have the outstanding opportunity to forge an
internationally collaborative activity to deliver a concrete
proposal on a five-year timescale for the implementation of an
infrastructure in which:
\begin{itemize}
  \item ENUBET and nuSTORM deliver the neutrino cross-section
    measurement programme identified in the ESPP and allow sensitive
    searches for physics beyond the Standard Model to be carried out;
    and
  \item A 6D muon ionisation cooling experiment is delivered as part
    of the technology development programme defined by the
    international Muon Collider collaboration.
\end{itemize}

\cleardoublepage
\pagenumbering{arabic}                   
\setcounter{page}{1}

\graphicspath{ {01-Introduction/Figures} }

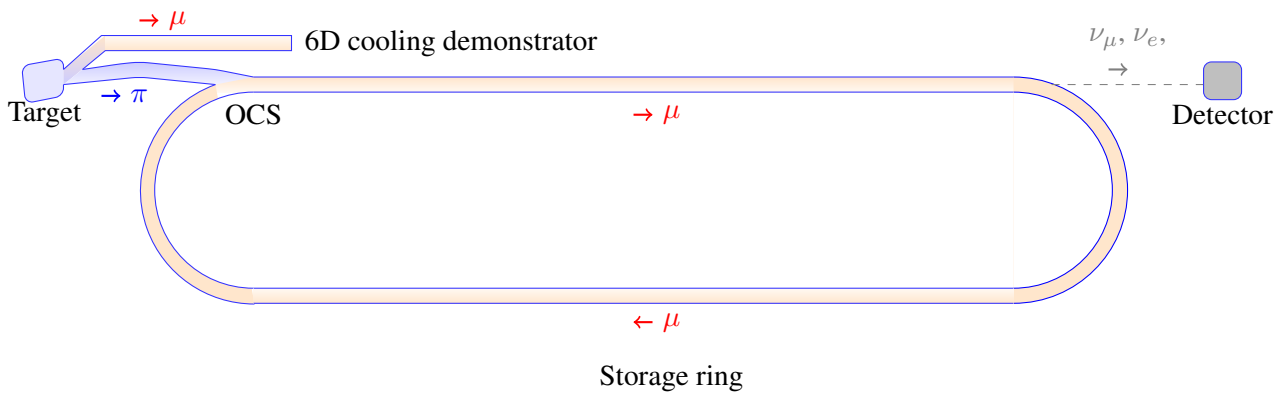
\begin{figure}[h]
\begin{tikzpicture}[scale=0.05]
\draw[blue!80, rounded corners=3pt, fill=blue!10] (10,-2) -- (10,-5) -- (0,-7) -- (-1,3) -- (10,5) -- (10, 2);
\node[] at (5,-10) {Target};

\shade[bottom color=blue!20,top color=orange!20] (10,-2) -- (21,7) -- (21,11) -- (10,2);
\shade[bottom color=orange!20,top color=orange!5] (70,11) -- (70,7) -- (21, 7) -- (21, 11);
\draw[blue!80] (10,2) -- (20,11) -- (70,11) -- (70,7) -- (21, 7) -- (15, 2);
\node[] at (112,9) {6D cooling demonstrator};
\draw[thick,->, color=red] (30,15) -- (35,15);
\node[text=red] at (40,15) {$\mu$};

\shade[top color=orange!20, bottom color=orange!5] (50, 0) -- (260,0) -- (260,-4) -- (50, -4);
\draw[blue!80, fill=orange!20] (50,-2) arc (110:270:30);
\draw[blue!80, fill=white] (60,-4) arc (90:270:26);
\draw[blue!80] (60, 0) -- (260,0);
\draw[blue!80] (60, -4) -- (260,-4);
\draw[blue, fill=orange!20] (260,-60) arc (-90:90:30);
\draw[blue, fill=white] (260,-56) arc (-90:90:26);
\shade[top color=orange!5, bottom color=orange!20] (60, -56) -- (260,-56) -- (260,-60) -- (60, -60);
\draw[blue!80] (60, -56) -- (260,-56);
\draw[blue!80] (60, -60) -- (260,-60);
\draw[thick,->, color=red] (160,-10) -- (165,-10);
\node[text=red] at (170,-10) {$\mu$};
\draw[thick,<-, color=red] (160,-65) -- (165,-65);
\node[text=red] at (170,-65) {$\mu$};
\node[] at (170,-80) {Storage ring};

\shade[bottom color=blue!5,top color=blue!20] (15, 2) -- (25,3.5) -- (30,4) -- (35,3.5) -- (50,2) -- (60,0) -- (50,-2) -- (35,-0.5) -- (30,0) -- (25,-0.5) -- (10, -2);
\draw[blue!80, rounded corners=3pt] (15, 2) -- (25,3.5) -- (30,4) -- (35,3.5) -- (50,2) -- (60,0);
\draw[blue!80, rounded corners=3pt] (10, -2) -- (25,-0.5) -- (30,0) -- (35,-0.5) -- (50,-2);
\draw[thick,->, color=blue] (20,-5) -- (25,-5);
\node[text=blue] at (30,-5) {$\pi$};
\node[] at (60,-10) {OCS};

\draw[gray, dashed] (270,-2) -- (310, -2);
\draw[blue!80, fill=gray!50, rounded corners=3pt] (310,-6) rectangle (320,4);
\draw[thick,->, color=gray] (285,2) -- (290,2);
\node[text=gray] at (290,10) {$\nu_{\mu}$, $\nu_{e}$, };
\node[] at (315,-10) {Detector};
\end{tikzpicture}
  \setlength{\belowcaptionskip}{-8pt}
  \caption{
    Schematic of the nuSTORM muon and neutrino-beam facility.
    The proton beam extracted from the FNAL Main Injector or the CERN
    SPS or PS strikes a target.
    Pions are collected using a horn and directed into conventional
    transfer lines that transport pions either to nuSTORM or a 6D
    muon ionisation cooling demonstration experiment.
    ENUBET, which requires pion and kaon beams, can be served using a
    third transfer line (not shown).
  }
  \label{Fig:Intro:nuSTORM:Schema}
\end{figure}

\section{Introduction}
\label{Sect:Intro}

nuSTORM, the `Neutrinos from Stored Muons' facility, has been designed
to provide intense neutrino beams with well-defined flavour
composition and energy spectrum.
By using neutrinos from the decay of muons confined within a storage
ring, a beam composed of equal fluxes of electron- and muon-neutrinos
can be created for which the energy spectrum can be calculated
precisely. 
According to current design considerations, it will be possible to
store muon beams with momentum from 1\,GeV/c to 6\,GeV/c and a
momentum acceptance of $\pm 16$\%.
Through its unique characteristics, the nuSTORM facility will have the
capability to: 
\begin{itemize}[noitemsep,topsep=0pt]
  \item Serve a definitive neutrino-nucleus scattering programme
    with uniquely well-characterised $\parenbar{\nu}_e$ and
    $\parenbar{\nu}_\mu$ beams;
  \item Allow searches for physics beyond the Standard Model and light
    sterile neutrinos with the exquisite sensitivity necessary to go
    beyond the reach of the FNAL Short Baseline Neutrino programme;
    and
  \item Provide the technology test-bed required for the development of
    muon beams capable of serving in a multi-TeV lepton-antilepton
    (muon) collider.  
\end{itemize}

nuSTORM is based on a low-energy muon decay ring (see 
figure~\ref{Fig:Intro:nuSTORM:Schema}).
Pions, produced in the bombardment of a target, are captured in a
magnetic channel.
The magnetic channel is designed to deliver a pion beam with
momentum $p_\pi$ and momentum spread $\sim \pm 10\%\,p_\pi$ to the 
muon decay ring.
The pion beam is injected into the production straight of the decay
ring.
Roughly half of the pions decay as the beam passes through the
production straight.
At the end of the straight, the return arc selects a muon beam of momentum $p_\mu < p_\pi$ and momentum spread 
$\sim \pm 16\%\,p_\mu$ that then circulates.
Undecayed pions and muons outside the momentum acceptance of the ring
are directed to a beam dump.
Pions from the target can also be directed to a decay channel in which
low-energy muons are collected and transported to a 6D
ionisation cooling experiment.
ENUBET can be served with pion and kaon beams in the same complex
through the addition of a third transfer line from the target
complex.

A detector placed on the axis of the nuSTORM production straight will
receive a bright flash of muon neutrinos from pion decay followed by a
series of pulses of muon and electron neutrinos from subsequent turns
of the muon beam.
Appropriate instrumentation in the decay ring and production
straight will be capable of determining the integrated neutrino flux
with a precision of $\lsim 1$\%.
The flavour composition of the neutrino beam from muon decay is
known and the neutrino-energy spectrum can be calculated precisely
using the Michel parameters and the optics of the muon decay ring.
The pion and muon momenta ($p_\pi$ and $p_\mu$) can be optimised to
measure $\parenbar{\nu}\!\!_e A$ and $\parenbar{\nu}\!\!_\mu A$ 
interactions with per-cent-level precision over the neutrino-energy
range $0.3 \lsim E_\nu \lsim 5.5$\,GeV and to search for light sterile 
neutrinos with excellent sensitivity.

\section{Motivation}
\label{Sect:Motivation}

The case for the nuSTORM facility rests on three themes:
\begin{enumerate}
  \item The uniquely well-defined neutrino beam generated in muon decay
    can be exploited to make detailed studies of neutrino-nucleus
    scattering over the neutrino-energy range of interest to present
    and future long- and short-baseline neutrino oscillation
    experiments. 
    The high-flux beams illuminating the detectors of future
    long-baseline experiments will allow the accumulation of very
    large data sets.
    Projections of the rate at which data will be collected in
    long-baseline experiments indicate that the statistical error will
    be reduced to the percent level by 2028--30.
    To optimise the discovery potential of such facilities requires
    that the systematic uncertainties be reduced to the percent level
    on a comparable timescale.
    This can be achieved by dedicated cross-section measurements by
    which to break the correlation between the cross-section and flux
    uncertainties and to reduce the overall systematic uncertainty to
    a level commensurate with the statistical and other systematic
    uncertainties in experiments such as Hyper-Kamiokande and DUNE.

    The nuSTORM $\parenbar{\nu} N$ scattering programme is no less
    important for the next generation of short-baseline experiments
    for which uncertainties in the magnitude and shape of backgrounds
    to the sterile-neutrino searches will become critically
    important.
    At nuSTORM, the flavour composition of the neutrino beam is known
    and its energy spectrum may be determined precisely using the
    storage-ring instrumentation.
    The precise knowledge of the neutrino flux combined with advanced
    detector techniques that are currently being developed will allow
    nuSTORM to provide the measurements necessary to maximise the
    sensitivity of the next generation of long- and short-baseline
    experiments.
  \item The nuSTORM neutrino beam instrumented with state-of-the art and magnetised
    near and far detectors, will allow searches for physics Beyond the Standard Model of unprecedented
    sensitivity to be carried out.
    The signal to background ratio for this combination is of order
    ten and is much larger than for other accelerator-based projects.
  \item The storage ring itself, and the muon beam it contains, can be
    used to carry out the R\&D programme required to implement the
    next step in the incremental development of muon accelerators for
    particle physics.
    Muon accelerators have been proposed as sources of intense,
    high-energy electron- and muon-neutrino beams at the Neutrino
    Factory \cite{Bandyopadhyay:2007kx,Choubey:2011zzq} and as the
    basis for multi-TeV $l^+ l^-$ collisions at the Muon Collider
    \cite{:1900cvd,Holmes:2010zz}.
    An incremental approach to the development of the facility has
    been outlined in \cite{Kaplan:2012zzb}, which has the potential
    for the elucidation of the physics of flavour at the Neutrino
    Factory and to provide multi-TeV $l^+ l^-$ collisions at the Muon
    Collider.
    nuSTORM would be the first neutrino-beam facility to be based on a
    stored muon beam and will provide a test-bed for the development
    of the technologies required for a multi-TeV Muon Collider and/or
    a Neutrino Factory.
\end{enumerate}
Just as the three legs of a tripod make it a uniquely stable platform,
the three individually-compelling themes that make up the case for
nuSTORM constitute a uniquely robust case for a facility that will
be at once immensely productive scientifically and seminal in the
creation of a new technique for particle physics.

\graphicspath{ {02-Motivation/02-01-Neutrino-scattering/Figures/} }

\subsection{Neutrino-nucleus scattering}
\label{SubSect:nuNScat}

nuSTORM will allow unprecedentedly precise studies of both elementary
processes and neutrino-nucleus scattering to be performed.
These prospects are not only interesting by themselves as a source of
information about the axial structure of nucleons and nuclei, but also
crucial to achieve the high-precision goals of neutrino oscillation
experiments~\cite{NuSTEC:2017hzk,Branca:2021vis}.
Indeed, near detectors help to reduce systematic uncertainties but do
not turn oscillation analysis into a mere rescaling because near and far detectors are not identical, have different efficiencies and are
illuminated by different neutrino fluxes.
In addition, oscillation probabilities depend on the neutrino energy
which is not known on an event-by-event basis but has to be
reconstructed. 
To minimise any bias in neutrino-energy reconstruction a realistic simulation of the interaction process is necessary.    

\subsubsection{Elementary processes}
In this context, by elementary processes one understands
neutrino-nucleon interactions, whose relevance is often
underestimated. The available information about them is scarce and comes mostly from old bubble chamber experiments.
These cross-sections could be measured directly using hydrogen or
deuterium targets or indirectly with the help of hydrogen-enriched
targets and subtraction techniques. Examples of these are the {\it solid hydrogen} concept~\cite{Duyang:2019prb}, where the (anti)neutrino proton interactions are obtained from the subtraction of events in plastic (CH$_2$) and graphite (C) targets and a high pressure TPC with
hydrogen-rich gases (such as CH$_4$), where the cross-section on
hydrogen would be extracted using {\it transverse kinematic
imbalance}~\cite{Lu:2015hea,Hamacher-Baumann:2020ogq}.
nuSTORM is the ideal place for such experiments because of the
precision that can be achieved.
The input for event generators would be highly valuable.
Furthermore, the availability of both muon and electron flavours of
neutrinos under similar experimental conditions would allow the
investigation of flavor-dependent features such as radiative corrections
and non-standard (BSM) interactions.  

\begin{figure}[h!]
  \begin{center}
    \includegraphics[width=0.49\textwidth]{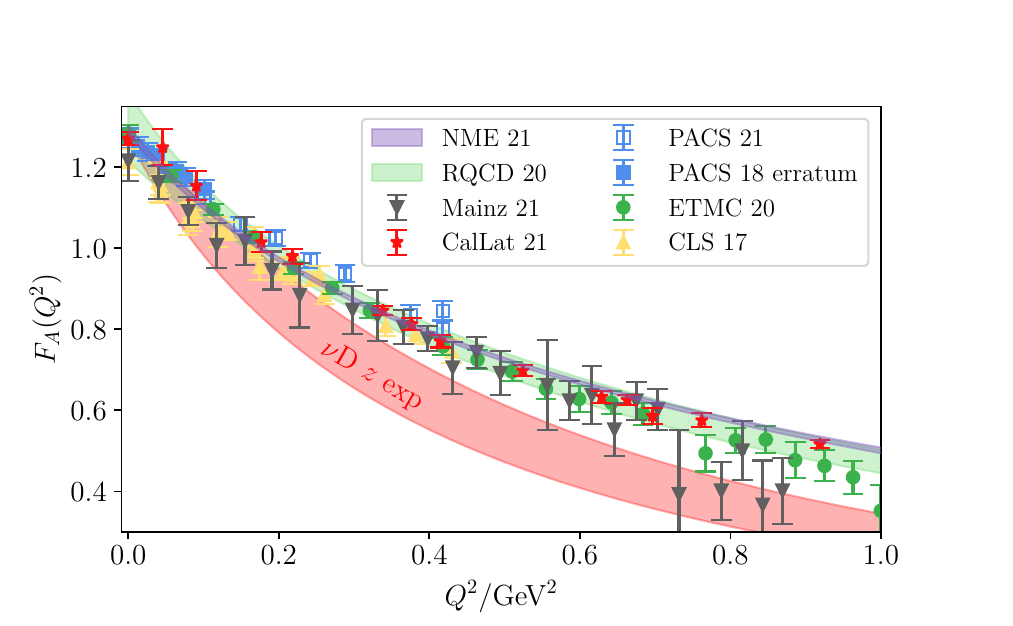}
    \includegraphics[width=0.49\textwidth]{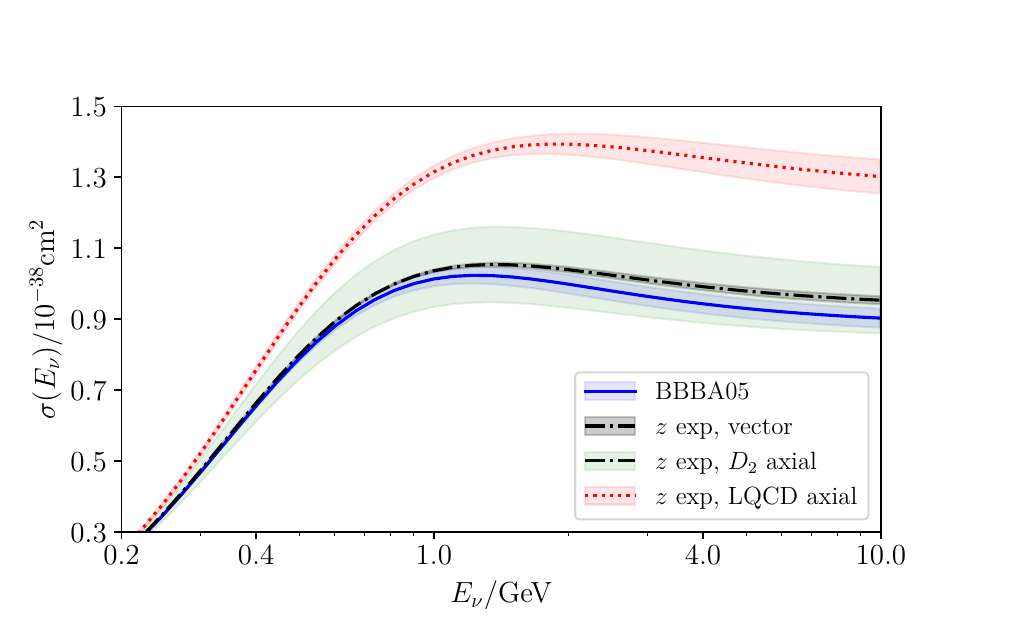}
  \end{center}
  \caption{Left: $F_A$ obtained in recent lattice-QCD studies compared to its determination from bubble chamber experiments using the z-expansion. Right: QE integrated cross-section from various parametrizations of the form factors. Details can be found in reference~\cite{Meyer:2022mix} from where these plots are taken.}
  \label{fig:QE}
\end{figure}

The simplest elementary process is charged-current quasielastic
scattering ($\nu_l \, n \rightarrow l^- \, p$ and
$\bar\nu_l \, p \rightarrow l^+ \, n$). 
Even for such a basic process, which could serve as a {\it standard
candle} to constrain neutrino fluxes, the dependence of the axial form
factor ($F_A$) on the four-momentum transferred to the nucleon squared
($Q^2$) is not precisely measured. 
Moreover, it has been noticed recently that lattice-QCD determinations
of $F_A(Q^2)$ are in fairly good agreement among themselves but in 
tension with empirical determinations~\cite{Meyer:2022mix} (see Fig.~\ref{fig:QE}, left, taken from this review). These lattice-QCD results would imply a 20\% increase of the quasielastic cross-section, as shown in the right panel of Fig.~\ref{fig:QE}, also from \cite{Meyer:2022mix}.  

Neutrinos also scatter inelastically on nucleons, predominantly
leading to single pion ($\pi N$) but also to $\gamma N$, $\pi \pi N$,
$\eta N$, $\rho N$, $K N$, $\pi \Sigma$, $\bar{K} N$, $K Y$, $\ldots$
final states.
For inelastic processes, the cross-section arises from the interplay
of resonant and non-resonant amplitudes, which become highly
non-trivial at higher invariant mass hadronic final states,
with several overlapping resonances  and coupled channels.
This is the shallow inelastic scattering region, where a large
fraction of events at DUNE will be found.
The dynamics have been investigated in detail in partial wave analyses
of large data sets available for photon, electron and pion-nucleon
interactions. This information is valuable to constrain weak inelastic processes and has  been used in their modelling as reviews, for instance, in reference~\cite{Nakamura:2016cnn}.  
However, the properties of the axial current at finite $Q^2$ remain
experimentally unconstrained. The transition from the resonant to the deep-inelastic scattering regime is also highly uncertain.
Quark-hadron duality on one hand and QCD  (higher-twist and target mass)
corrections on the other are valuable tools to describe it (see
reference~\cite{SajjadAthar:2020nvy} for a recent review), but
progress in their development is hindered by the lack of experimental
nuclear-effect-free information from elementary targets.
With muon momenta in the range $1 \leq p_\mu \leq 6$~GeV$/c$, the
resulting neutrino spectrum goes up to $\sim 5.5$~GeV 
and would make the detailed study of this region possible.

\subsubsection{Neutrino-nucleus interactions}
There is considerable interest in the study of neutrino scattering
on the heavy targets used in oscillation experiments.
nuSTORM will have a strong impact by characterising the flavour
differences which are particularly important at low energy and 
momentum transfers (in the laboratory frame).
These differences can arise from a subtle interplay between lepton
kinematic factors and response functions~\cite{Pandey:2016jju}.
The search for CP-invariance violation in present and planned
long-baseline neutrino-oscillation experiments is based on the
measurement of the rate of $\nu_e$ appearance in $\nu_\mu$ beams.
nuSTORM has the potential to perform high-statistics measurements of
the $\parenbar{\nu}_e$ cross-sections and, in particular, the
$\sigma(\nu_e)/\sigma(\nu_\mu)$ cross-section ratio, which is among
the largest systematic uncertainties at DUNE~\cite{DUNE:2015lol}.
With the help of nuSTORM,  the required sensitivity to CP violation
can be reached with a smaller exposure.  

Measurements of quasielastic-like scattering at nuSTORM can also lead
to a better description of initial state nucleon-nucleon correlations
and meson-exchange currents, which are known to provide a sizeable
contribution to the semi-inclusive electron scattering cross-section
and have been found important at MiniBooNE and T2K: comparisons of
different theoretical results to data can be found, for example, in
Figures~8-9 of reference~\cite{Wilkinson:2016wmz} (MiniBooNE) and
in Figures~7-9 of reference~\cite{T2K:2016jor} (T2K).
The comparisons of the SUSA model to these data have been recently
summarised in reference~\cite{Barbaro:2021psv}.
Discrepancies with theory (or, at least with its generator
implementation) have been found at the higher energy and momentum
transfers probed at MINERvA and NOvA as can be appreciated in
references~\cite{MINERvA:2015ydy,NOvA:2020rbg}.
With unprecedented understanding of the beam flux (see
section~\ref{sec:flux}) and sophisticated detector designs (see
section~\ref{SubSect:DetectorConsiderations}), nuSTORM can play an
important role in understanding these differences.

The characterisation of nuclear corrections to parton distribution
functions will also benefit from precise measurements of the inclusive
neutrino-nucleus cross-section, to unravel the differences in nuclear
effects observed in weak and electromagnetic processes and to resolve
the tensions that have been observed by nCTEQ. It was suggested that $\nu  A$ and $l^\pm A$ data could only be reconciled if the correlations in $\nu  A$ were not taken into account. However, a more recent  comprehensive nCTEQ analysis indicates that neglecting correlations does not relieve the tension between $\nu  A$
and $l^\pm A$ data. More precise data on a wider variety of nuclear targets would be most welcome. 

\begin{figure}[htp]
  \begin{center}
    \includegraphics[width=0.45\textwidth]{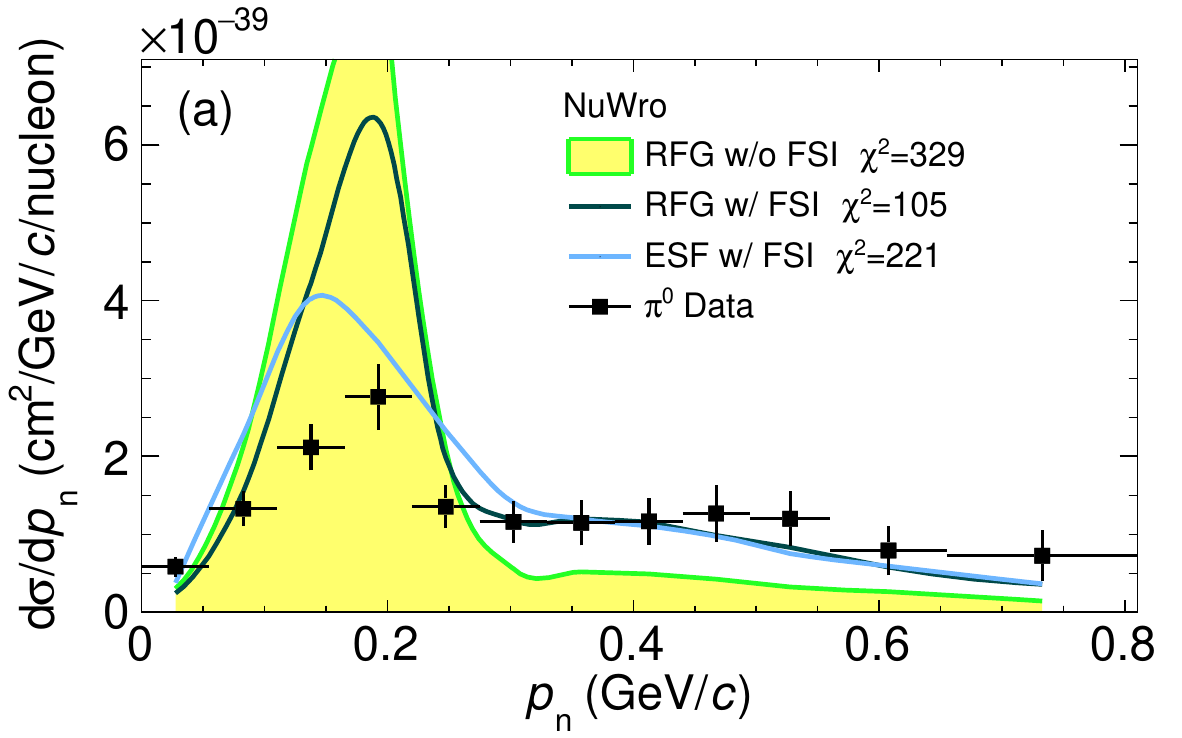}
    \includegraphics[width=0.45\textwidth]{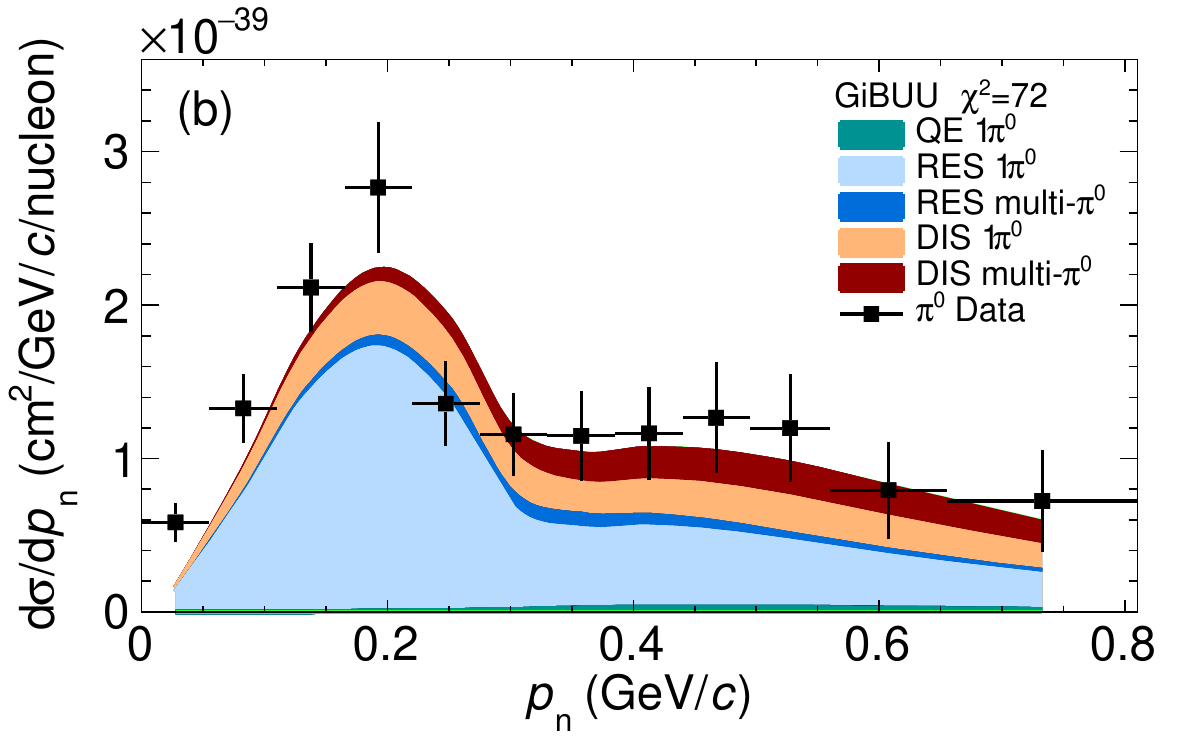}
  \end{center}
  \caption{
    Flux-averaged differential cross-section in the emulated nucleon
    momentum, $p_\textrm{n}$, measured by MINERvA with the Low-Energy
    NuMI beam flux peaking at 3~GeV on a scintillator (CH) target.
    The peak region of the $p_\textrm{n}$ distribution directly
    reflects the Fermi motion of the struck neutron in the
    charged-current $\pi^0$
    production~\cite{Lu:2015tcr,Furmanski:2016wqo,Lu:2019nmf}.
    Comparisons with NuWro~\cite{Golan:2012wx} and
    GiBUU~\cite{Leitner:2006ww,Leitner:2008ue,Buss:2011mx} predictions
    are made.
    Figures from reference~\cite{MINERvA:2020anu}.
  }
  \label{fig:minervapizerotki}
\end{figure}

With a suitable detector set, nuSTORM can also study exclusive
channels in neutrino-nucleus scattering (see
section~\ref{sec:simevent}). These include one and two-nucleon knockout but also single and multiple meson production. These reactions are largely influenced by strong final state interactions between the produced particles and the nuclear
environment. Pions, in particular, can scatter, change charge or be absorbed on
their way out of the nucleus~\cite{Mosel:2019vhx}.
Pion production will play an important role in the future neutrino
oscillation programme.
Pioneering measurements of pion production by MINERvA (cf. review
article~\cite{MINERvA:2021csy} and references therein) have shown
tensions with model predictions (see, for example,
figure~\ref{fig:minervapizerotki}).
Accurate modelling of these interactions are crucial to reduce biases
in calorimetric neutrino energy determination. 
\graphicspath{ {02-Motivation/02-02-BSM/Figures/} }

\subsection{Searches for physics beyond the Standard Model}
\label{SubSect:BSM}

\subsubsection{Rare scattering processes}
If the precision of the nuSTORM concept is combined with high
statistics, the study of {\it rare} processes with small cross-sections becomes feasible.
Processes in this category include neutrino-electron scattering,
coherent meson production, weak and electromagnetic production of 
single photons and dileptons.
In addition to elementary quasielastic scattering, neutrino-electron
scattering or even coherent meson production could be precisely
measured and used as {\it standard candles} for flux determination in
other experiments.
Furthermore, weak couplings and $\sin^2{\theta_W}$ can be extracted,
providing a precision test of the Standard Model (SM).
Unlike DUNE, nuSTORM has access to both $\nu_\mu-$ and
$\nu_e-$electron scattering.      

While these exotic processes are allowed in the SM, precision
measurements can disclose physics beyond the Standard Model (BSM).
This is the case in neutrino tridents in which neutrino scattering off
the Coulomb field of a heavy nucleus generates a pair of charged
leptons.
The existence of light $Z'$ or other particles in the dark sector can
modify the trident cross-section~\cite{Altmannshofer:2014pba}.

Single photon emission in neutral current interactions is another rare
process that has received attention as a background in $\nu_e$
appearance measurements in Cherenkov detectors.
Its cross-section has never been measured and, so far, only upper
limits from NOMAD, T2K and, more recently, MicroBooNE are available.
Besides, some of the proposed explanations of the MiniBooNE anomaly
involve the production of a heavy (1--100~MeV) neutrino via
electromagnetic ($\gamma$ mediator), weak ($Z$) or BSM ($Z'$)
interactions, leading to a signal in the single photon or $e^+e^-$
channels (see reference~\cite{Alvarez-Ruso:2021dna} for a recent
review).
While recent MicroBooNE results disfavour some explanations of the
MiniBooNE anomaly, the full range of possible solutions is still
unexplored~\cite{Abdallah:2022grs}. 

\subsubsection{Short-baseline flavour transitions and sterile neutrino
  searches}

The unique neutrino beam composition at nuSTORM allows to use $\mu^+ \rightarrow e^+ \, \nu_e
\bar\nu_\mu$ decay to search for short-baseline oscillations using muon final states. In particular, for the first time, nuSTORM would allow to search for $\nu_\mu$ appearance from $\nu_e
\rightarrow \nu_\mu$ oscillations, which is not subject to photon-like and intrinsic $\nu_e$ contamination as in other accelerator experiments. This appearance measurement relies on a good charge identification to discriminate the $\mu^-$ signal from the intrinsic $\mu^+$ background from the muon-decay 
$\bar\nu_\mu$ component. The latter can also be used to search for $\bar\nu_\mu \rightarrow \bar\nu_\mu$ disappearance. The low flux systematics allow to reach greater levels of precision in the total normalisation of the event rate, and one can expect this measurement to be limited by the cross-section uncertainties.
Nevertheless, greater sensitivity can be achieved by identifying spectral distortions of the
$\mu^+$ spectrum in the detector; this would require accurate momentum
measurement.

In Ref.~\cite{nuSTORM:2014phr}, a detailed study of the sensitivity of the previous Fermilab design of nuSTORM to sterile neutrinos was performed. This was based on a far detector located at $2$~km from the muon storage ring and a total of $10^{21}$~POT, corresponding to approximately $2\times10^{18}$ useful muon decays. The detector was a $1.3$~kt magnetised iron-scintillator detector with excellent muon charge discrimination. The final sensitivity was found to be greater than $5\sigma$ throughout the entire region of oscillation parameter space preferred by the MiniBooNE and LSND results. Ref.~\cite{Ahdida:2020whw} expanded the scope of the oscillation search to show that nuSTORM can also provide very stringent tests of the unitarity of the neutrino mixing matrix, non-standard interactions, as well as Lorentz and CPT symmetries.

In summary, nuSTORM would be capable of addressing open questions
concerning the non-unitarity of the neutrino mixing matrix, non-standard
interactions, Lorentz invariance (and CPT violation) and provide a
definitive test of light sterile
neutrinos~\cite{nuSTORM:2014phr,Ahdida:2020whw}.

\graphicspath{ {02-Motivation/02-03-RnD/Figures/} }

\subsection{Technology test-bed}
\label{SubSect:RnD}

A Muon Collider has the potential to deliver lepton-anti-lepton
collisions at centre-of-mass energies up to 10\,TeV at a cost and on
a timescale advantageous when compared to electron-positron or
next-generation hadron colliders~\cite{Adolphsen:2022bay}.
The international Muon Collider collaboration is developing conceptual
designs for facilities capable of operation at centre-of-mass
energies of 3\,TeV and 10\,TeV~\cite{iMC:WWW}.

nuSTORM will have the world's highest power stored muon beam.
Such a beam will provide the opportunity to develop and test
technologies that will be critical to the delivery of muon beams with the
brightness necessary for the Muon Collider to deliver the specified
luminosity of $\sim 10 \times 10^{34}\,{\rm cm}^{-2}\,{\rm s}^{-1}$.
In particular, a low-energy muon beam may be produced through the
capture of an appropriate pion-beam phase space at the nuSTORM target.
The muon beam derived from pion decay could be directed towards a
muon ionisation cooling system designed to demonstrate the feasibility of
reducing the muon-beam phase space in all six phase-space dimensions. 

The principle of ionisation cooling was demonstrated by the MICE
collaboration~\cite{MICE:Nature}.
The MICE experiment determined the change in the transverse emittance
of a muon beam as it passed through a single liquid-hydrogen or
lithium-hydride absorber.
The muon beams used by MICE had momentum in the range 140\,MeV/c to
240\,MeV/c and emittance in the range 3\,mm to 10\,mm.
For the Muon Collider to achieve its design luminosity requires
transverse and longitudinal emittances of $25\,\mu$m and 7.5\,MeV\,m
respectively.
An experiment based on the lessons learnt at MICE, can be used to develop
the techniques to compress the 6D phase-space volume of a muon beam
to the values required to achieve the luminosity specification of the 
Muon Collider.

\subsubsection{Muon ionisation cooling}
Ionisation cooling is effected by passing a muon beam through a
material (the absorber), in which it loses energy, and subsequently
accelerating the beam to restore the energy lost in the absorber.
The ionisation cooling process occurs on a short timescale and
is therefore able to cool the beam efficiently with modest decay
losses.
The net effect of the energy-loss/re-acceleration process is to
reduce the transverse beam size, i.e. to reduce the
\emph{transverse emittance}.

Multiple Coulomb scattering is detrimental to the cooling
performance.
Energy absorbers made from materials having low atomic number, such as
lithium or hydrogen, are preferred as the ratio of ionisation energy
loss to multiple Coulomb scattering is favourable.
If the beam is tightly focused the effect of multiple Coulomb
scattering relative to the intrinsic beam divergence is reduced.
Therefore ionisation cooling lattices must be designed to maintain
very tight focusing. 

Random fluctuations in energy loss, known as energy straggling, tend
to result in an increase in the beam-energy spread and hence
longitudinal emittance.
It is possible to reduce the longitudinal emittance of the beam by
introducing a dipole and wedge-shaped absorber onto the beamline.
The dipole bends lower momentum particles more strongly, introducing a
correlation between energy and position.
By aligning the thicker part of the wedge with high energy particles
and the thinner part of the wedge with low energy particles, the
energy spread is removed and replaced with an increased spread in
position.
This is an emittance exchange process; the energy spread of the beam
is decreased at the expense of an increased transverse emittance.
Transverse emittance, in turn, is reduced by the ionisation cooling
process.
Overall the size of the beam in the 6D phase space $(x, p_x, y, p_y,
t, E)$ is reduced. 

\subsubsection{6D cooling system design}
The baseline design for the Muon Collider ionisation cooling system is
of order 1\,km long.
The main part comprises the rectilinear cooling system described in
\cite{PhysRevSTAB.18.031003}.
In this system, focusing is achieved by means of solenoids having
fields up to around 12\,T.
A weak dipole field is introduced that creates a position-energy
correlation and wedge-shaped absorbers are used to deliver 6D
cooling.
Early parts of the system, where the beam has large emittance, are
optimised for large acceptance, while later parts of the system are
optimised for cooling to low emittance. 

\begin{figure}[htp]
  \begin{center}
    \includegraphics[width=\textwidth]{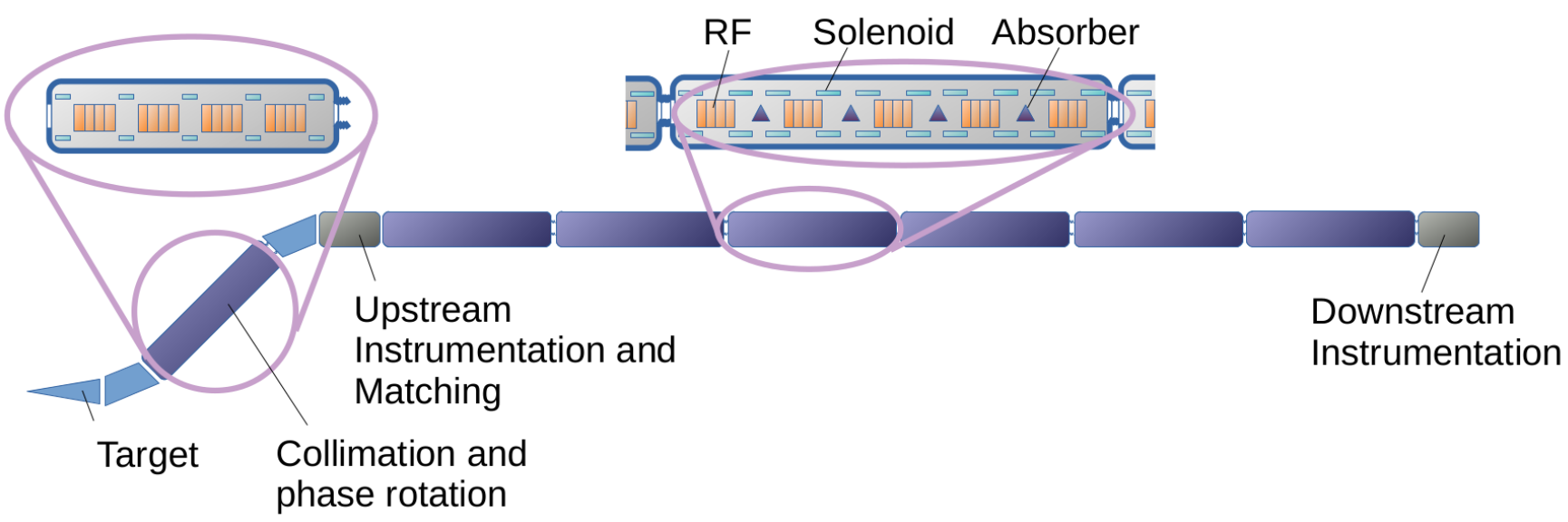}
  \end{center}
  \caption{
    Schematic of a possible implementation of the Muon Collider
    demonstrator for 6D cooling.
  }
  \label{fig:cooling_schematic}
\end{figure}

The proposed demonstrator facility would comprise around 50\,m of
cooling equipment, as shown in figure~\ref{fig:cooling_schematic}.
Low momentum pions are diverted from the target region by means of a
dipole switchyard.
Off-momentum pions are rejected and the resultant beam is collimated
while the pions decay to muons.
Collimation is necessary in order to deliver a low emittance beam
suitable for demonstration of later stages of the cooling system.
A short, high voltage section of RF then accelerates or decelerates
particles that are out of phase with the RF cavities yielding a
bunched beam.
Finally, a short focusing system is used to ensure parameters such as
muon-beam divergence and position spread are matched to the cooling
channel focusing system.  

For beam cooling, a tightly packed lattice of RF cavities, absorbers
and solenoids is envisaged.
The later stages of the rectilinear cooling system operate in the
second stability region, where each cell of the magnet lattice has two
foci; one at the absorber and a weaker focus in the centre of the RF
cavities.
This can be achieved with solenoids having fields up to about 10\,T.
Sufficient dispersion can be achieved by appropriate choice of dipole
polarity and field, so that low angle wedges can be used while
yielding a satisfactory longitudinal cooling performance. 

Instrumentation is required upstream and downstream of the cooling
system to measure the beam emittance and a dedicated section of
beamline is foreseen to support this.
Additional instrumentation is required in each module to support
operations such as beam alignment and RF phasing.
Beam intersecting devices or instrumentation of the absorbers may be
an effective means of monitoring the muon beam, less readily achievable in 
a more conventional beamline.

\section{nuSTORM facility; overview}
\label{Sect:nuSTORM_overview}

The following sections give a brief description of the development of
the nuSTORM concept since it was presented in~\cite{Ahdida:2020whw}.
The authors are actively developing the accelerator design, the
detector concept, and the analysis framework.
Therefore the summary presented below should be considered a snapshot
in the development of the nuSTORM facility.

\graphicspath { {03-nuSTORM-facility/03-01-Accel/Figures/} }
\subsection{Accelerator facility}
\label{SubSect:AccelFacility}

\subsubsection{Target and pion transport}
\label{SubSect:Accel-facility:targetandtransfer}

The feasibility of implementing nuSTORM at the SPS at CERN was
presented in~\cite{Ahdida:2020whw}. 
The proton beam extracted from the SPS at 100\,GeV is focused on a
solid (low-$Z$) target placed inside a focusing horn.
In the simulations presented below, the proton  beam impinges on an
inconel target.
Other materials for the target, such as graphite, may be
considered.
Particles emerging from the target are focused by the horn and
collected in a short transfer line with a large momentum acceptance of
$\sim \pm 10\%$.
The transfer line is composed of dipoles, collimators and quadrupoles.
It is proposed that the target, horn and the initial part of the
transfer line are contained in an inert helium atmosphere to reduce
activation and corrosion of beam-line equipment by limiting the
presence of ozone and nitrogen oxides.
The target-and-collection system will be installed underground in a
cavern, with a shaft giving access to a surface building.
The shaft and surface buildings will be offset with respect to the
incoming proton beam direction, the target, and the outgoing pion
beam.

The design of the pion transfer line is based on the initial FNAL
design.
The design was modified to accommodate the projected radiation hazards
and an improved injection scheme.
A modular construction scheme, allowing for a greater degree of
flexibility during the design phase and using simple quadrupole FODO 
cells and achromatic dipole bends, was adopted. 
An initial capture section will be present inside the initial
containment vessel, which will be followed by the proton absorber.
A series of collimators will be used in addition to the bending
sections to reduce the radiation load on the downstream beam lines. 

The first achromatic bending section is used to divert particles
within the desired momentum range away from the proton absorber
towards the ring. 
This is key to reduce the radiation dose to downstream elements and
provide a momentum selection for the transmitted pion beam. 
A quadrupole FODO lattice is used to transport the beam to a second
achromatic bending section followed by beta-function matching and
injection into the ring. 
The length of this section was chosen to ensure the radiation
contamination within the arc sections of the ring is minimised, whilst
being short enough to ensure that pions of low momentum can be
successfully transported. 
Building on~\cite{Ahdida:2020whw}, the beta-function and dispersion
matching section has been defined and the first integration of the
transfer line with the ring injection system has been performed.

\begin{figure}
  \begin{center}
    \includegraphics[width=0.95\textwidth]{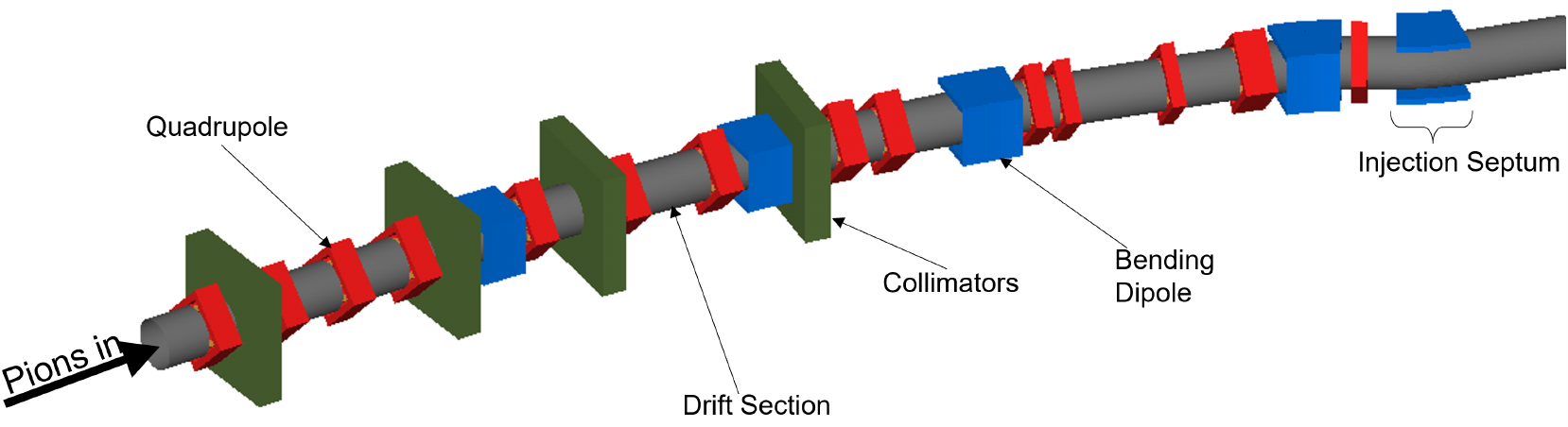}
  \end{center}
  \caption{
    Schematic drawing of the pion transfer line generated in BDSIM code.
  }
  \label{Fig:Accel-facility:targetandtransfer:line}
\end{figure}

\begin{figure}
  \begin{center}
    \includegraphics[width=0.85\textwidth]{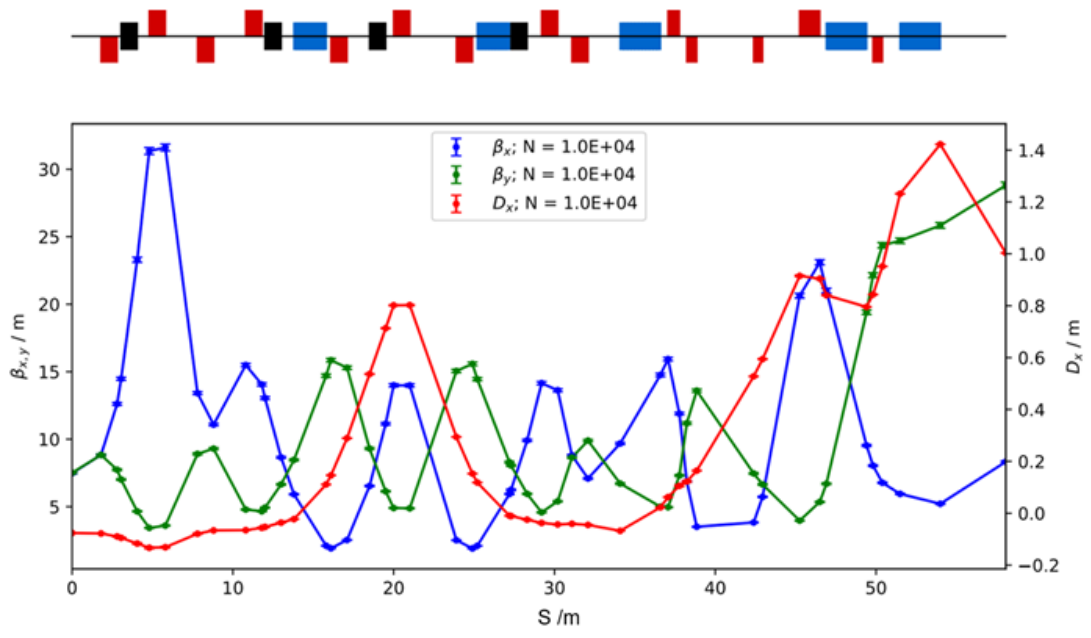}
  \end{center}
  \caption{
    The $\beta$ functions ($\beta_x$- blue, $\beta_y$- green) and
    dispersion ($D_x$- red) in the pion transfer line downstream from
    the horn until the end of the injection septum generated using
    beam dynamics simulations in BDSIM code.
  }
  \label{Fig:Accel-facility:targetandtransfer:optics}
\end{figure}

Pion production in the target was simulated using MARS
\cite{Mokhov:2017klc} and FLUKA \cite{FLUKA,FLUKA2}.
The particle-distribution was used as input for beam-dynamics studies
of the transfer line using BDSIM code~\cite{Nevay:2021wtg}. 
The beam dynamics simulations confirmed the large momentum acceptance
of the transfer line.
The layout of the transfer line simulated in BDSIM is shown in
figure~\ref{Fig:Accel-facility:targetandtransfer:line}.
The optical functions of the transfer line, shown in
figure~\ref{Fig:Accel-facility:targetandtransfer:optics}, were
reproduced using beam dynamics simulations in BDSIM.
The values of the optical functions obtained with BDSIM agree well
with the design values and provide the correct beam conditions at the
injection to the ring. 

\subsubsection{Storage ring design}
\label{SubSect:Accel-facility:muRng:muStrRng}

\begin{figure}
  \begin{center}
    \includegraphics[width=0.95\textwidth]{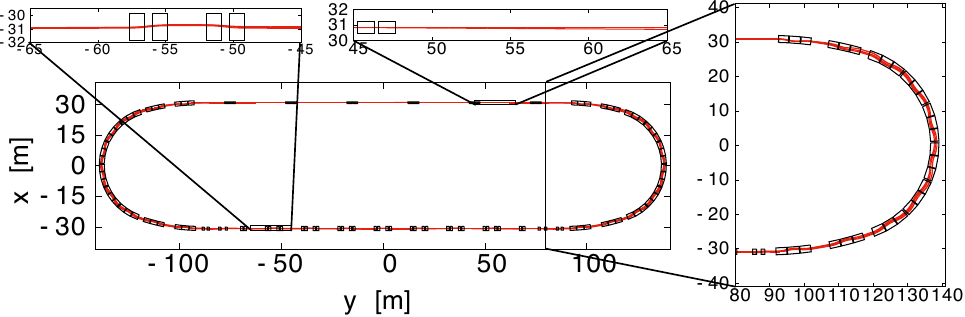}
  \end{center}
  \caption{
    Schematic drawing of the revision of the muon storage ring.
    The beam circulates in an clockwise direction.
    The production straight (at $x\sim 30$\,m) is composed of large
    aperture quadrupoles that produce the large values of the betatron
    function required to minimise the divergence of the neutrino beam
    produced in muon decay.
    The lattices of the arcs and return straight are based on the
    Fixed Field Alternating gradient (FFA) concept and allow a large dynamic
    aperture to be maintained. 
  }
  \label{Fig:Accel-facility:muRng:muStrgRng}
\end{figure}

The nuSTORM decay ring, shown schematically in
figure~\ref{Fig:Accel-facility:muRng:muStrgRng}, is a compact racetrack
storage ring with a circumference of $\sim 616$\,m that incorporates
large aperture magnets.  
In order to include the orbit combination section (OCS), used for
the stochastic injection of the pion beam into the ring, a dispersion
suppressor is needed between the arc and the production straight. 
Strong bending magnets are also needed in the arcs to minimise the arc
length, in order to maximise the number of useful muon decays.

\begin{figure}
  \begin{center}
    \includegraphics[width=0.65\textwidth]{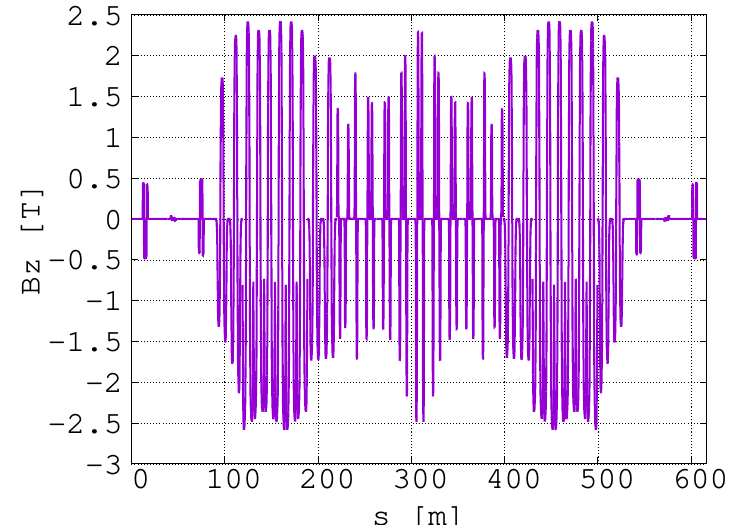}
  \end{center}
  \caption{
    The vertical magnetic field for the maximum momentum ($\sim
    6$\,GeV/c) muon closed orbit in the racetrack FFA ring.
  }
  \label{Fig:Accel-facility:muRng:muStrgRngB}
\end{figure}

Several designs for the nuSTORM storage ring have already been
proposed based either on a separated function magnet or Fixed Field
Alternating gradient (FFA) approach~\cite{Adey:2013pio,Liu:2017sya, Lagrange:2018triplet}.
To serve the neutrino-scattering programme, the ring was redesigned to  
store muon beams with a momentum of between 1\,GeV/c and 6\,GeV/c with
a momentum acceptance of up to $\pm 16$\%, thereby increasing the
neutrino flux. To keep the momentum acceptance and transverse dynamic
acceptance large, and simultaneously to maximise the muon accumulation
efficiency, a hybrid concept was developed
(figure~\ref{Fig:Accel-facility:muRng:muStrgRng}). 
Conventional FODO optics, used in the production straight, are
combined with FFA cells, for which the chromaticity is zero, in the
arcs and in the return straight. 
This allows the revised lattice to achieve:
\begin{itemize}[noitemsep,topsep=0pt]
  \item Zero dispersion in the quadrupole injection/production
    straight; 
  \item Zero chromaticity in the arcs and in the return straight,
    thereby limiting the overall chromaticity of the ring; and thus 
  \item Large overall transverse and momentum acceptance.
\end{itemize}
The arcs exploit superconducting combined-function magnets with  
magnetic fields of up to $\sim 2.6$\,T.
The return straight is based on combined-function room-temperature
magnets. 
The production straight uses large-aperture room temperature
quadrupoles. The vertical magnetic field around the ring for the
maximum momentum ($\sim 6$\,GeV/c) muon closed orbit in the racetrack
FFA ring is shown in figure~\ref{Fig:Accel-facility:muRng:muStrgRngB}.
The mean betatron functions in both the production and return 
straights are kept large enough to minimise the contribution of 
betatron oscillations to the angular spread of the neutrino beam, 
such that both can be used to serve a neutrino-physics programme.

\begin{figure}
  \begin{center}
    \includegraphics[width=0.75\textwidth]{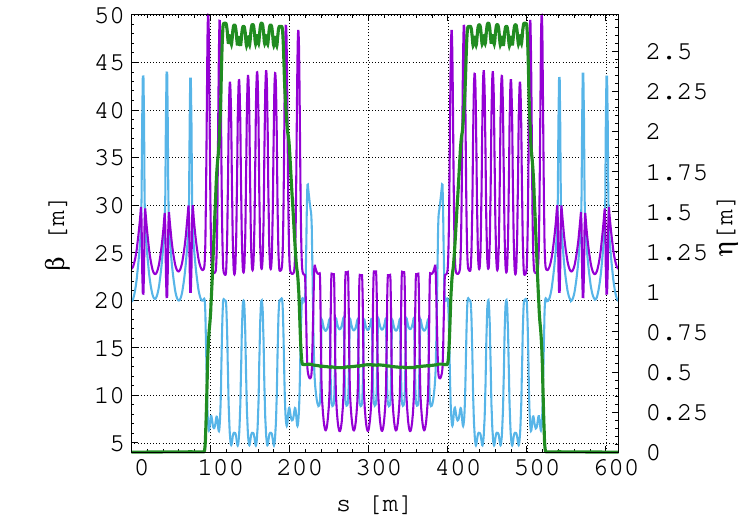}
  \end{center}
  \caption{
    The betatron functions (horizontal-blue and vertical-purple) and
    dispersion (green) for reference momentum ($5.2$\,GeV/c) muon
    closed orbit in the racetrack FFA ring.
  }
  \label{Fig:Accel-facility:muRng:muStrgRngTwiss}
\end{figure}

\begin{figure}
  \begin{center}
    \includegraphics[width=0.65\textwidth]{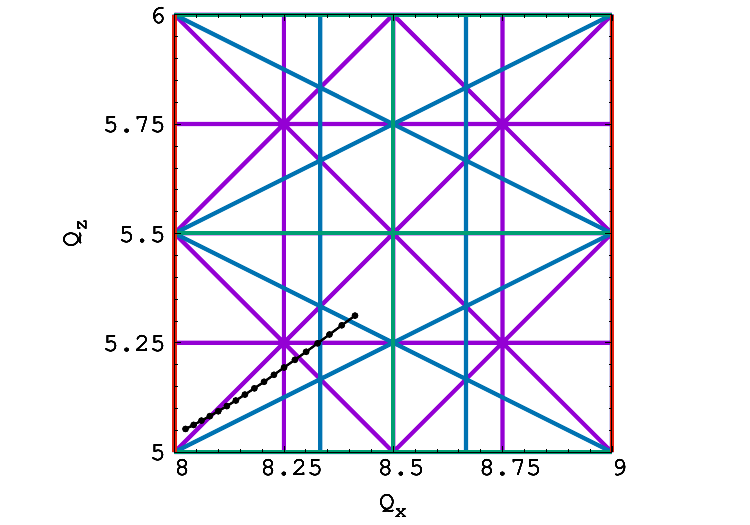}
  \end{center}
  \caption{
    The machine tunes for the muon beam stored in the nuSTORM ring at
    the reference momentum of 5.2\,GeV/c with the momentum spread of
    $\pm 16$\%.
  }
  \label{Fig:Accel-facility:muRng:muStrgRngTune}
\end{figure}

The arc cells have a high magnet-packing factor to minimise the 
arc length and are connected with the injection and return straights
using specific matching sections.
The matching section serving the injection straight matches dispersion
to zero and allows a long straight for injection to be accommodated.
Additional matching sections are between the arcs and the cells of the
return straight. 
The Twiss parameters around the ring are shown in
figure~\ref{Fig:Accel-facility:muRng:muStrgRngTwiss}.
Selected parameters of the hybrid design for the racetrack ring are
summarised in table~\ref{tab:Accel-facility:muRng:muStrgRngP}.
The reference tunes of the machine ($8.203$, $5.159$) are chosen such
that they are not close to the dangerous resonances.
The off-momentum tunes have been chosen to avoid integer and
half-integer resonances (see
figure~\ref{Fig:Accel-facility:muRng:muStrgRngTune}).
Further reduction of the chromaticity of the ring is possible by
altering the nonlinear magnetic field distribution in the regular arc
cells.

\begin{table}[!h]
    \centering
\begin{tabular}{lcc}
\hline
\hline
Total circumference & 616~m\\
Length of one straight section & 180~m\\
One straight section/circumference ratio & 29\%\\
Operational momentum range & 1--6~GeV/c\\
Reference momentum & 5.2~GeV/c\\
Reference tunes ($Q_h$, $Q_V$) & ($8.203$, $5.159$)\\
Momentum acceptance & $\pm 16\%$\\
\hline
Number of cells in the ring: &\\
Straight quad cells & 6 \\
Arc first matching cells & 4\\
Arc cells & 12\\
Arc second matching cells & 4\\
Straight matching FFA cells & 1 (+1 mirror) \\
Straight FFA cells & 8 \\
\hline
\hline
\end{tabular} 
 \caption{Selected parameters of the hybrid FFA storage ring.}
 \label{tab:Accel-facility:muRng:muStrgRngP}
\end{table}

\begin{figure}
  \begin{center}
    \includegraphics[width=0.45\textwidth]{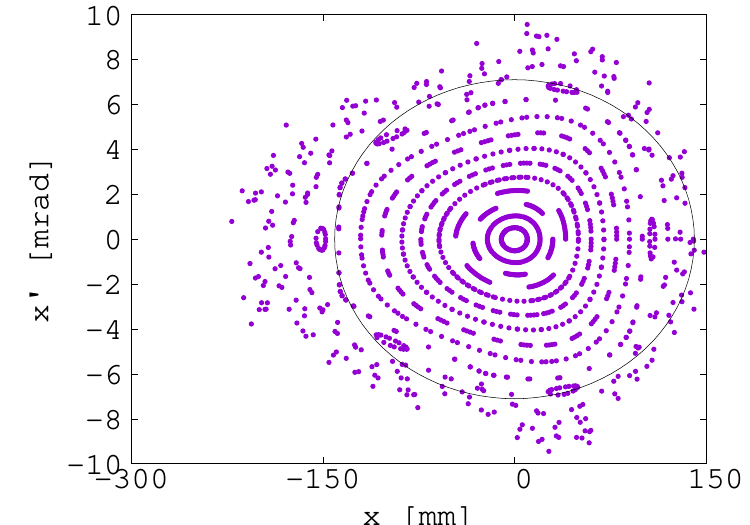}
    \includegraphics[width=0.45\textwidth]{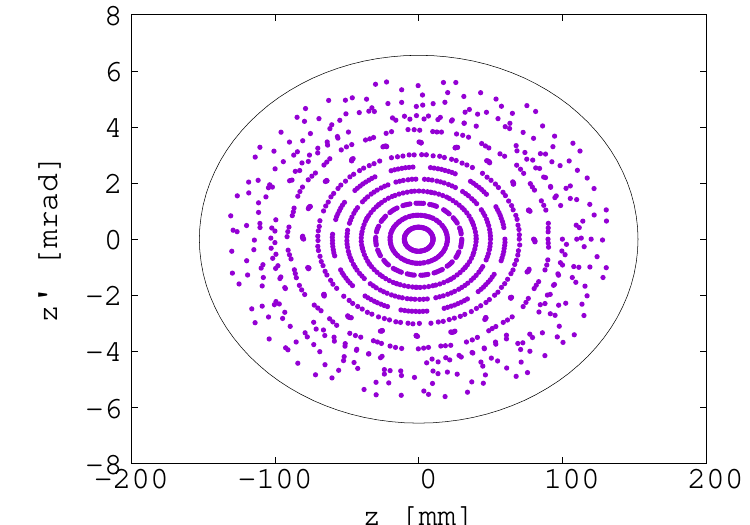}
  \end{center}
  \caption{
    The horizontal (left-hand plot) and the vertical (right-hand plot)
    dynamical acceptance studies in the hybrid nuSTORM ring at the
    reference momentum of 5.2\,GeV/c.
    Particles are tracked over 100 turns with different amplitudes in
    the plane of study including a small off-set from the closed orbit
    in the other plane.
    The black ellipse represents the acceptance of $1\,\pi$\,mm\,rad.
  }
  \label{Fig:Accel-facility:muRng:muStrgRngA}
\end{figure}

The performance of the hybrid FFA design for the storage ring was
verified in tracking studies.
In order to incorporate tracking through the combined-function magnets,
taking into account the fringe fields and large amplitude effects, a
code used for the full FFA machine developed previously was
used~\cite{ Lagrange:2018triplet}.
It is a stepwise tracking code based on Runge-Kutta integration, using
Enge-type fringe fields.
The results of the multi-turn tracking show that the dynamical
acceptance of the machine is about $1\,\pi$\,mm\,rad in both transverse
planes, which is required for the needs of the experimental
programme, as shown in figure~\ref{Fig:Accel-facility:muRng:muStrgRngA}.
The studies to cross-check the results with the PyZgoubi code, as
performed successfully before~\cite{ Lagrange:2018triplet}, are
underway.

\graphicspath{ {03-nuSTORM-facility/03-02-DetectorConsiderations/Figures/} }

\subsection{Detector considerations}
\label{SubSect:DetectorConsiderations}

As has been described, the detector at the nuSTORM facility will
deliver a rich programme of neutrino-interaction physics that can be
explored with unprecedented precision and an unprecedented reach in
searches for new phenomena.
In order to take full advantage of these opportunities, the detector
requirements extend far beyond that needed for a 3-flavor
oscillation search.
Options for the detectors are discussed below, but we list some of the
overarching performance requirements here: 
\begin{itemize}
\item Highly segmented detectors capable of operation at high event rate.  Detectors with precise 3D tracking (or very precise timing) capability over 4$\pi$  are required.
\item Detectors with excellent muon and electron ID capability.
\item Detectors with excellent energy resolution. 
\item A magnetised detector so that the charge of the muon and electron in the final state can be determined.  In addition reconstruction via spectrometry can be applied to event reconstruction as opposed to being done via calorimetry.  This is particularly important for higher energy nuSTORM tunes neutrino interactions where the outgoing muon's momentum must be measured via spectrometry.
\item Detectors with excellent hadronic particle ID, i.e. p/$\pi$/K separation at momenta from a few hundred MeV/c to a few GeV/c.
\item Detectors with neutron detection capability (with energy determination).
\item A detector that presents a variety of nuclear targets to measure cross-sections as a function of the nuclear target mass number A.
\end{itemize}
Many of the detector concepts now incorporated in the upgraded T2K near detector~\cite{Blanchet:2021pqb} and those being developed for DUNE~\cite{DUNE:2021tad} are appropriate for detectors at nuSTORM.  These concepts include:
\begin{enumerate}
    \item Highly-segmented tracking scintillator detector (SuperFGD);
    \item Pixelated LAr detector;
    \item Magnetised high-pressure gaseous Ar TPC (HPgTPC);
    \item Straw-Tube trackers (STT) with thin targets.
\end{enumerate}
Magnetisation of all these detectors is under consideration.
The SuperFGD for T2K is a magnetised detector as will be the STT for
DUNE.
The HPgTPC is by design a magnetised detector.
Although magnetisation concepts for a pixelated LAr have been
developed, the high cost for the magnet system presents obstacles to
its use, although R\&D on high-temperature superconductor and cable
may make this option affordable.

\begin{figure}[h]
  \label{fig:ND-GAr}
  \begin{center}
    \includegraphics[width=0.7\textwidth]{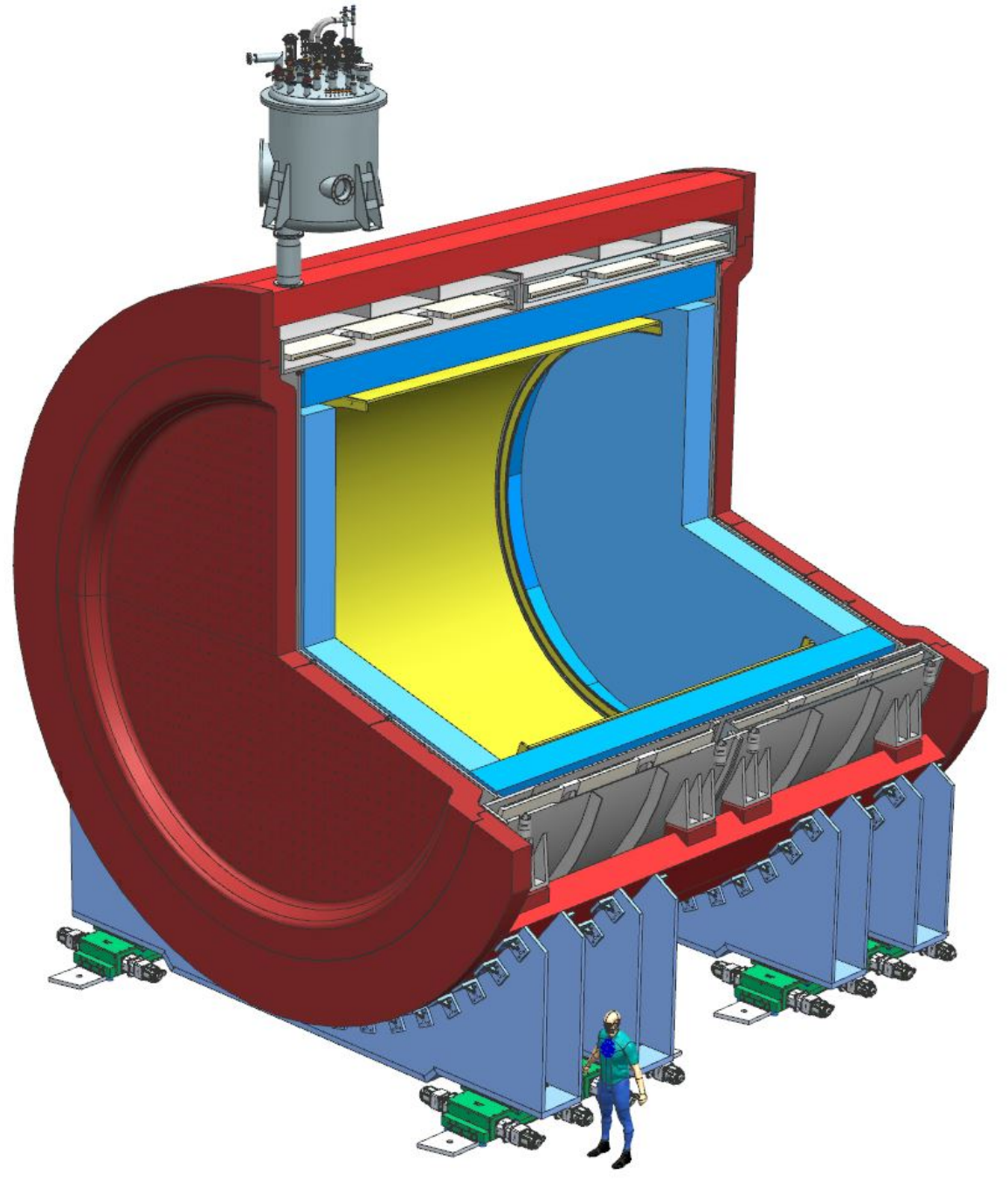}
  \end{center}
  \caption{
    Schematic of the ND-GAr concept.
    From inside out, Yellow: HPgTPC, Blue: ECAL, Grey: Superconducting
    solenoid in its cryostat and Red: Return Iron.
  }
\end{figure}

Although the concept of magnetisation in neutrino detectors is not
new, the application of a collider-detector design for neutrino
physics is.
One such example is the high-pressure gas TPC (HPgTPC) detector
concept (called ND-GAr) for the DUNE near detector
complex~\cite{Bersani:2021blp}.
An overview of the detector is shown in Figure~\ref{fig:ND-GAr}.
ND-GAr is a large detector with a magnetic volume that is
approximately 7\,m in diameter and 7.5\,m long (both the HPgTPC and the
ECAL are in the magnetic volume).
The solenoid magnet and the return iron provide pressure containment.
This detector offers many advantages including: capability to vary the
target nucleus (main gas component) from He to Xe, operation at
pressures from 1 Bar to 10 Bar, 4$\pi$ tracking with track thresholds
down to 5 MeV, excellent particle ID which allows for very precise
determination of exclusive final states and the addition of a magnetic
field allows for energy measurement via spectrometry as well as
calorimetry (from the ECAL).
In DUNE, ND-GAr functions as muon catcher for the pixelated LAr
detector which is just upstream.
The return iron has a window which allows muons that exit the LAr to
be accurately momentum analysed in the HPgTPC. 

\subsection{Neutrino fluxes}\label{sec:flux}

We consider the neutrino energy spectrum at the front face of a
detector of area 5\,m by 5\,m placed 50\,m beyond the end of the nuSTORM production straight. We present the results for two pion energies, 3~GeV and 5~GeV and three neutrino signals: $\nu_{\mu}$ from pion decay in the production straight, referred to as pion flash; $\nu_{\mu}$ and $\nu_{e}$ from muon decays in time with the pion decays in the production straight; $\nu_{\mu}$ and $\nu_{e}$ from muons which decay in the production straight of the nuSTORM ring after the end of the pion flash. These numbers for the 5~GeV pion beam are normalised to the number of protons on target; for the 3~GeV beam there are still unexplained losses in the FLUKA simulation and so we show results normalised to the number of pions accepted by the transfer line. The number of pions produced when modelling the production target with MARS and FLUKA are sufficiently different that attempting to normalise rates to protons on target would be misleading.The results are presented assuming a 100 GeV proton beam on the production target; we have not considered in detail the effect of using the 26 GeV beam from the CERN PS. The distributions are similar in pion angle and energy, but the 100 GeV beam produces about a factor of 5 more pions at all energies. At present we are agnostic on the choice of primary beam energy, merely noting that a lower energy requires more intensity or longer running for the same sensitivity. 

Neutrinos from muons which decay before they reach the end of the
production straight and are captured by the ring,  overlap in time
with the pion decays in the production straight and constitute a
background to the pion flash signal of around 1\%.
While neutrinos from muons which are captured by the nuSTORM ring and
decay during subsequent  rotations round the ring, are essentially
background free.
A number of other sources of backgrounds are considered and where it
can be shown that the contamination will be well below 1\%, no attempt
is made to simulate them in detail. A description of the simulation and
background estimation can be found in Appendix \ref{Appen:Simulation} and \ref{Appen:BkgEstimates}.

\subsubsection{Pion flash neutrinos at $E_{\pi}$=5 GeV and 3 GeV}
\begin{figure}
  \centerline{
  \includegraphics[scale=0.45]{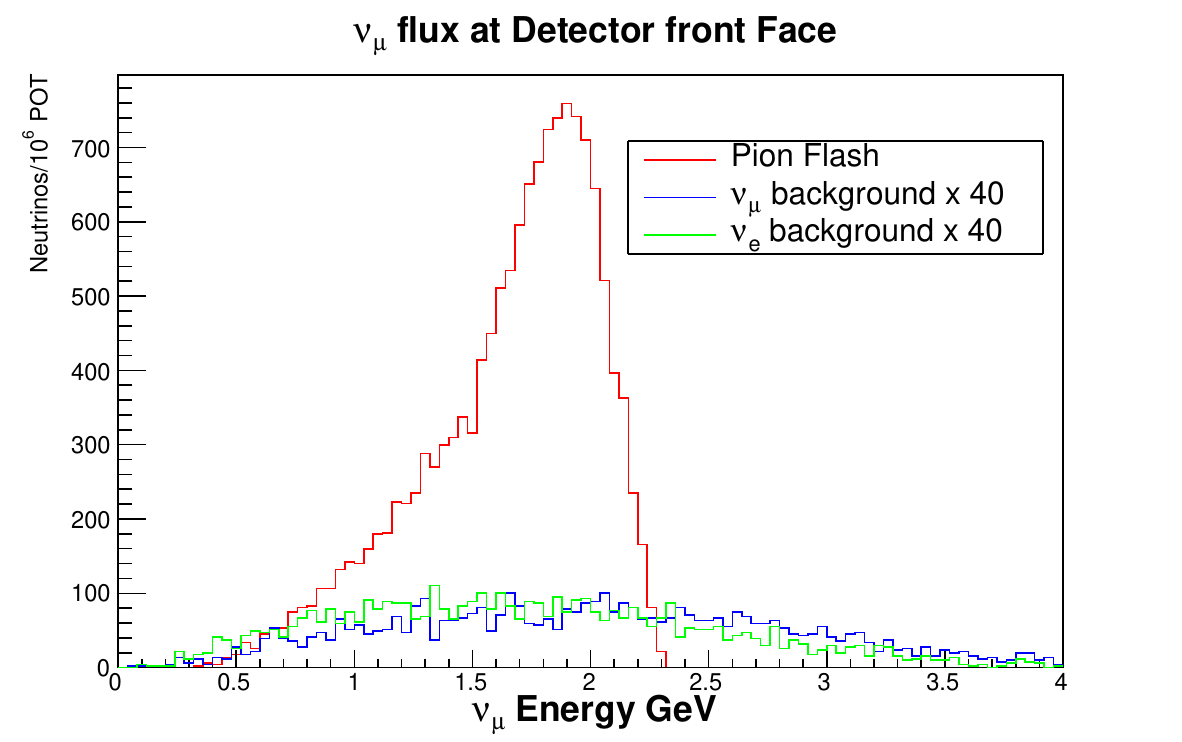}
  \includegraphics[scale=0.45]{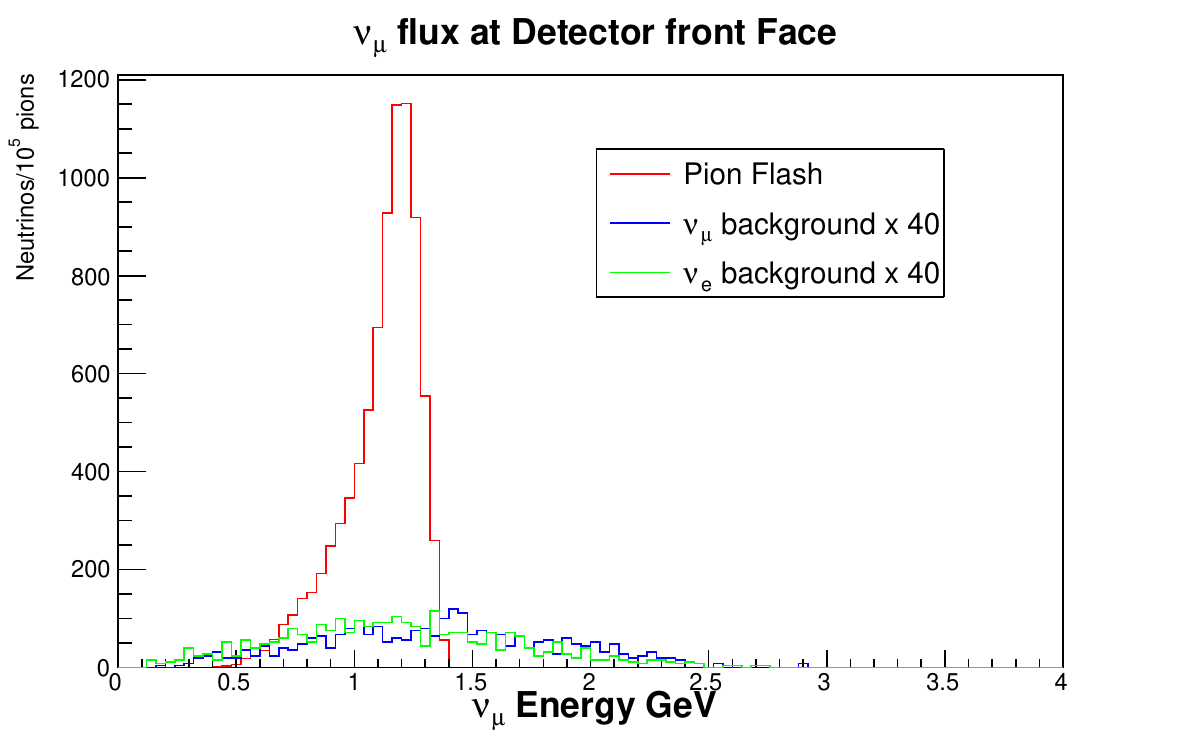}
  }
  \caption{
    Left: Pion flash energy spectrum for $E_{\pi}$=5~GeV  and
    associated muon backgrounds.
    Right: Pion flash energy spectrum for $E_{\pi}$=3~GeV and
    associated muon backgrounds.
  }
  \label{fig:EPiFlash}
\end{figure}

The biggest source of neutrinos are those from the pion flash.
Figure \ref{fig:EPiFlash} shows the energy spectrum of those pions.
The plot on the left is for a central $E_{\pi}$ of 5 GeV and right a
central $E_{\pi}$ of 3 GeV, in each case with a $\pm 10\%$ momentum
bite, corresponding to the design parameters of the machine.
The background from in-time muon decays is shown scaled up by a factor
40. 

\subsubsection{Muon signal neutrinos $E_{\pi}$=5 GeV and 3 GeV}

The number of $\nu_{\mu}$'s and $\nu_{e}$'s, which reach the front
face of the detector is similar; their energy spectrum is similar with
the  $\nu_{\mu}$'s being slightly harder.
We simulated three times as many events at 3 GeV in order to give us a
comparable number of events at 3 GeV and 5 GeV.
Dropping the central pion energy to 2 GeV looses another factor of 2.
This energy dependence is largely due to the way the angular
distribution of the neutrinos broadens as the $Q$ value of the decay
becomes a larger fraction of the beam's kinetic energy.
The distance of the detector front face from the end of the production
straight has not been optimised, but when we start detailed design of
the hall and accelerator layout, it will be important to keep this
distance as short as possible.

\begin{figure}
  \centerline{
  \includegraphics[scale=0.45]{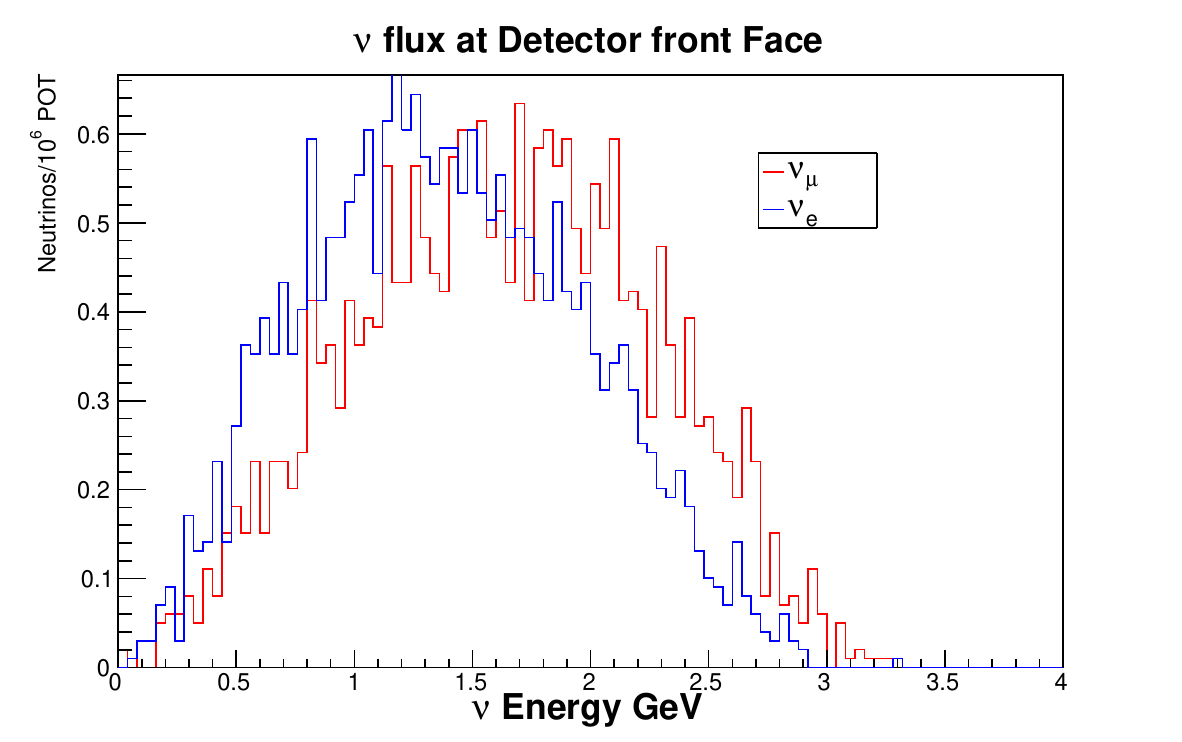}
  \includegraphics[scale=0.45]{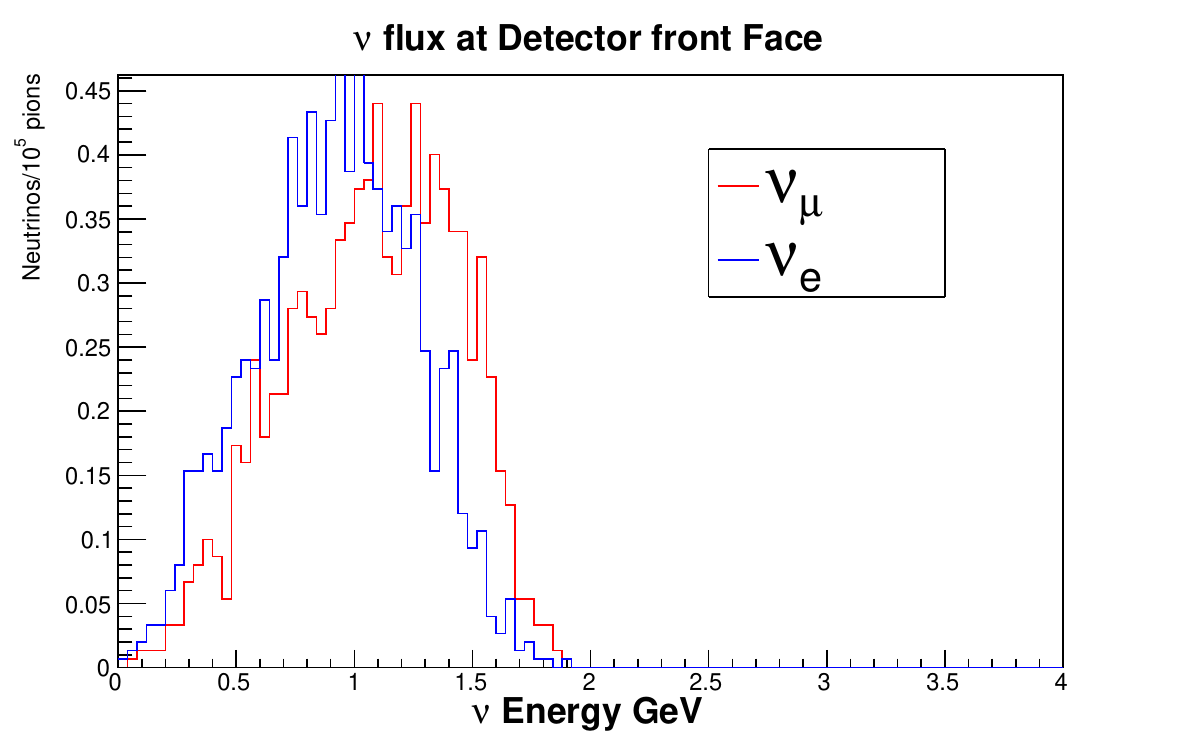}
  }
  \caption{
    Signal neutrinos from muon decay.
    Left: $\nu_{\mu}$  (Red) and $\nu_{e}$ (Blue) at $E_{\pi}$=5~GeV
    normalised to protons on target.
    Right:  $\nu_{\mu}$  (Red) and $\nu_{e}$ (Blue) at $E_{\pi}$=3~GeV
    normalised to pions entering the transfer line.
  }
  \label{fig:E5MuSig}
\end{figure}

\subsection{Event composition and kinematic distributions}\label{sec:simevent}
Measurements of neutrino interaction cross-sections on various target
nuclei has been a significant experimental focus in the last 10-15
years of neutrino physics as their importance to the systematics
budget of the long baseline neutrino oscillation experiments became
clear.
In recent years the T2K experiment has produced data on $\nu_{\mu}$ and $\nu_{e}$
interactions on scintillator at an average neutrino energy of 600~MeV\cite{T2K:2020txr,T2K:2020jav,Jenkins:2021ggh}, and
the $\mbox{MINER}{\nu}\mbox{A}$\cite{MINERvA:2021csy,MINERvA:2020anu,MINERvA:2019ope,MINERvA:2018hba} experiment has produced similar data but at average neutrino energies of
3~GeV and on a variety of target nuclei. The results of both experiments show
that, although the primary lepton kinematics are reasonably well predicted by our models, the
hadronic multiplicities, identities and kinematics are still poorly described. A detector for nuSTORM meeting the overarching requirements described above will address these issues.

In order to quantify the detector requirements,
GENIE~v3.06\cite{Andreopoulos:2009rq} was used to simulate neutrino
interactions on a carbon nuclear target. As input to this simulation, 
the $\nu_{\mu}$ flux for an initial pion energy of 5~GeV and 3~GeV
discussed above were implemented.
As the fluxes are not, presently, absolutely normalised, only the
relative event category composition can be studied.
The shapes of the kinematic distributions of final state particles
which might be visible in a detector can, however, be used to inform
the detector design. 

The relative event rate composition for interactions of neutrinos from muon decay in the production straight is shown in table~\ref{table:EventRateComposition}
for the 5~GeV and 3~GeV pion beams.
\begin{table}[h]
\centering
\begin{tabular}{|l|c|c|} \hline
$\nu_{\mu}$ Channel  & {$E_{\pi} = 3$~GeV} & {$E_{\pi} = 5$~GeV} \\ \hline
CC Quasi-elastic     & 0.41 & 0.32\\
2p2h                 & 0.12 & 0.12\\
Resonance Production & 0.43 & 0.49 \\
Deep Inelastic & 0.03 & 0.09 \\
Coherent Pion  & 0.005  & 0.005 \\ \hline
\end{tabular}
\caption{ Relative event category composition of $\nu_{\mu}$ interactions in a Carbon target. Shown are the relative event rates using $\nu_{\mu}$
from muon decay in the production straights, for two energies of initial pions from the target.}
\label{table:EventRateComposition}
\end{table}
For either pion energy, the event sample is dominated by the
quasi-elastic and resonance interaction channels.
The transition from quasi-elastic to resonant meson production is
known as the dual region and is poorly understood.
A large sample of neutrino interaction in this region would be crucial
to understanding the physics.
In addition, there is a smaller component of the Deep Inelastic
Scattering (DIS) interaction channel.
This is, however, in a low $Q^2$ and $W^2$ region, known as the
Shallow Inelastic region.
This region, which represents the transition from resonant meson
production to the DIS region, is not well-understood, neither
experimentally nor theoretically.

\begin{figure}[h]
  \begin{center}
    \mbox{\subfigure{\includegraphics[width=8.0cm]{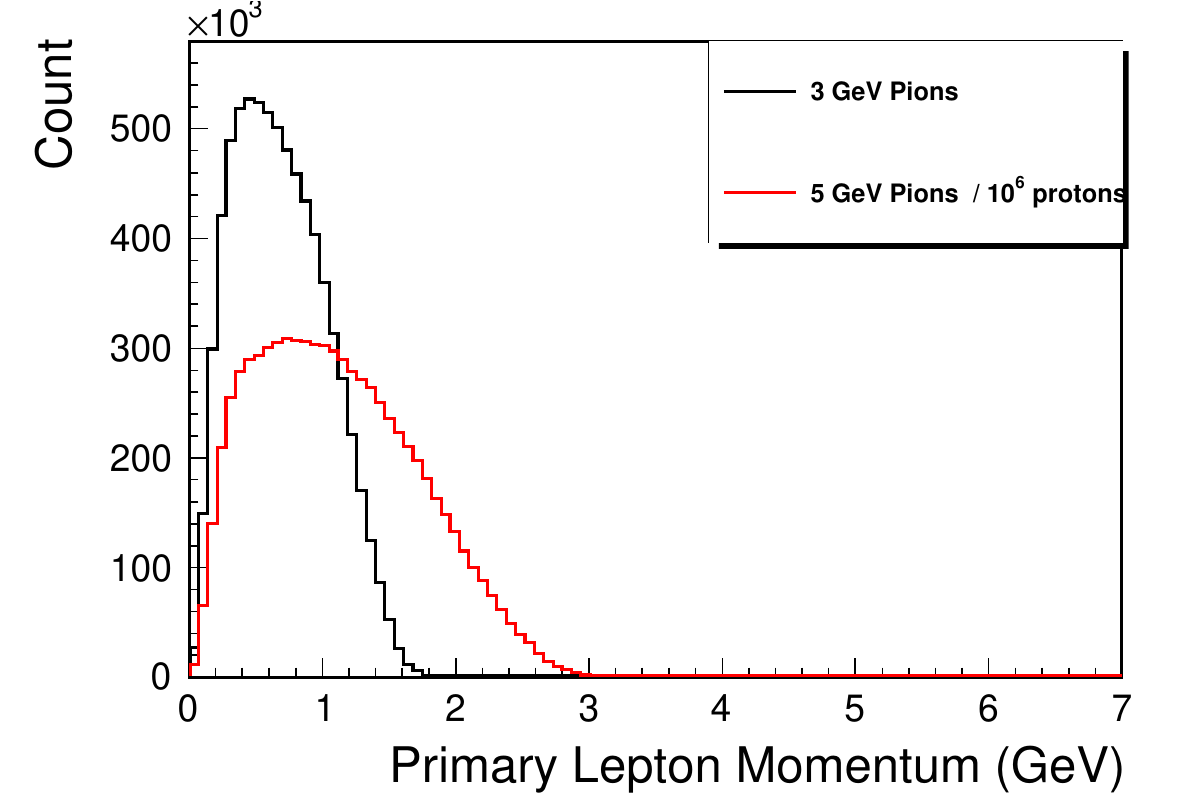}}\quad
          \subfigure{\includegraphics[width=8.0cm]{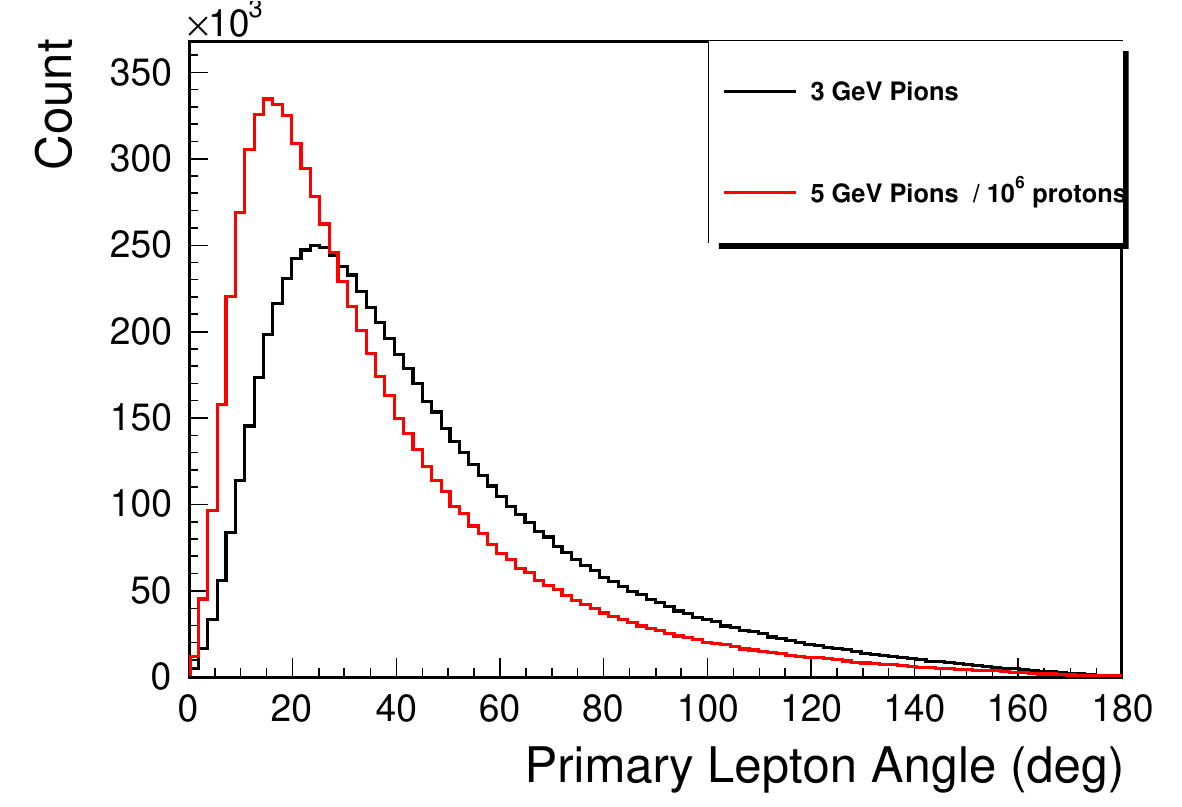} }}
  \end{center}
  \caption{
    (Left) Momentum of the primary muon from $\nu_{\mu}$ interactions
    on carbon. The neutrinos are generated from muon decay in the
    nuSTORM production straight and are generated with two different
    energies for the pions at the target. (Right) Angle of the primary
    muon from $\nu_{\mu}$ interactions on carbon with respect to the
    beam direction.
  }
  \label{fig:PrimaryLeptonKinematics}
\end{figure}

\begin{figure}[h]
  \begin{center}
    \includegraphics[width=8.0cm]{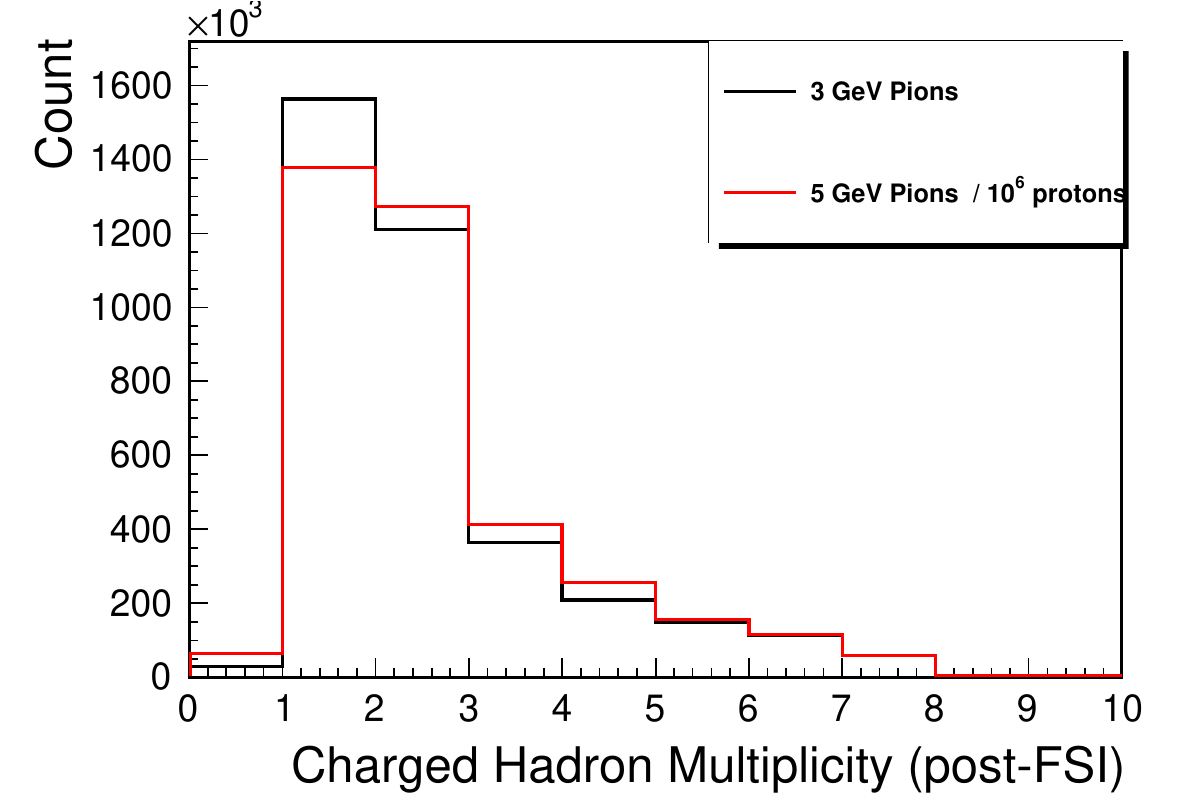}
  \end{center}
  \caption{
    Charged hadron multiplicity in the detector.
  }
  \label{fig:HadronMultiplicity}
\end{figure}

\begin{figure}[h]
  \begin{center}
    \mbox{\subfigure{\includegraphics[width=8.0cm]{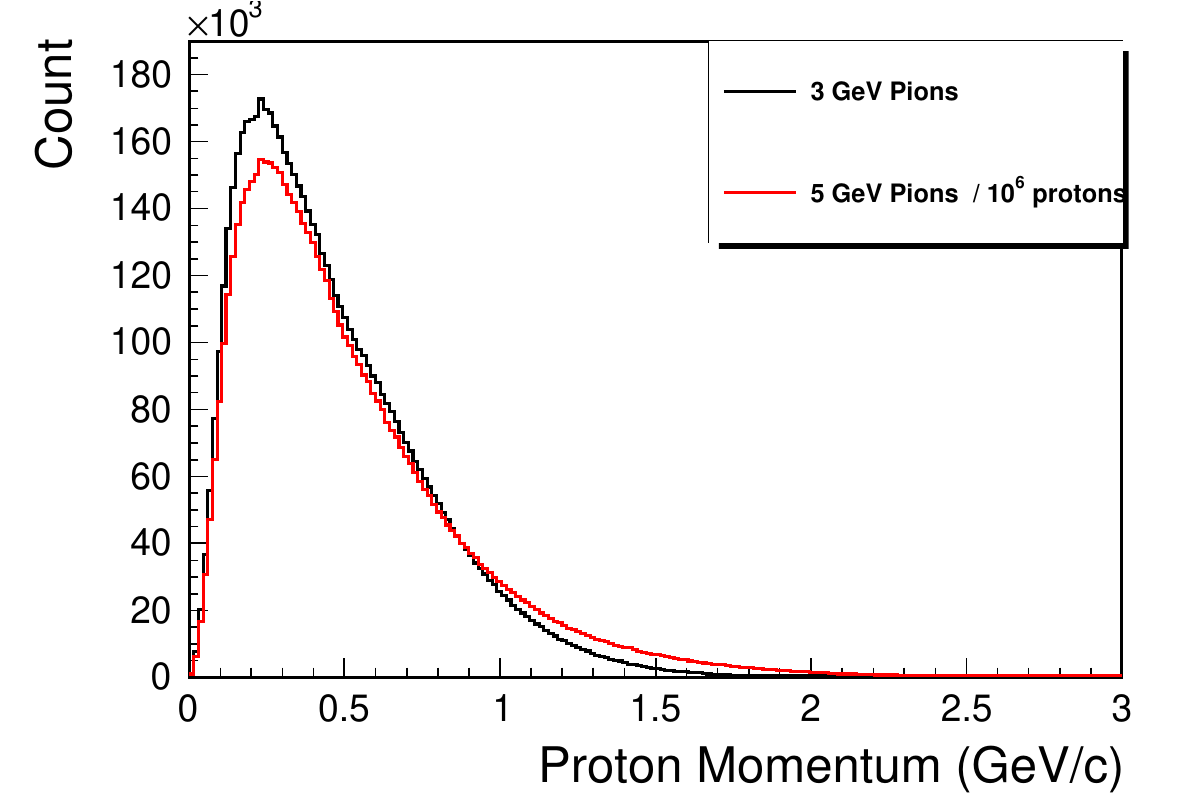}}\quad
          \subfigure{\includegraphics[width=8.0cm]{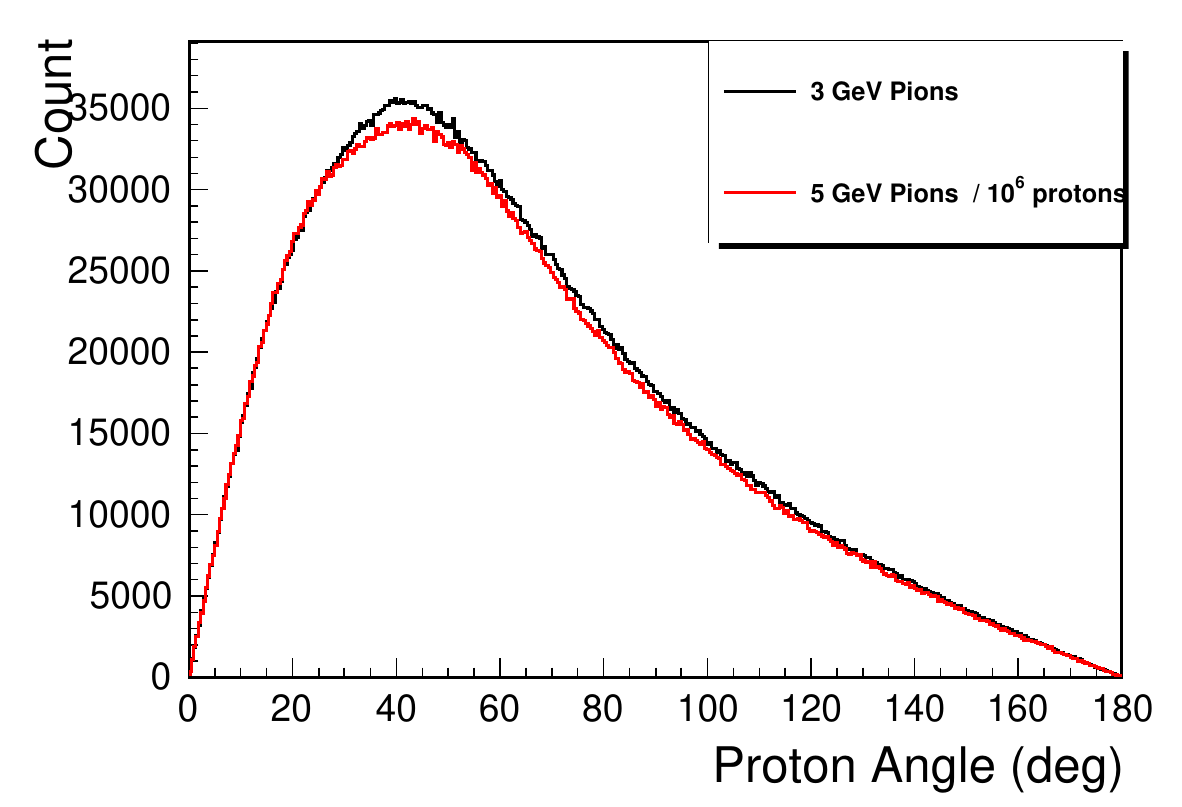}}}
  \end{center}
  \caption{
    (Left) Momentum of protons in the hadronic final state of
    $\nu_{\mu}$ interactions on carbon.
    The neutrinos are generated from muon decay in the nuSTORM production
    straight and are generated with two different energies for the pions
    at the target.
    (Right) Angle of protons in the hadronic final state from
    $\nu_{\mu}$ interactions on carbon with respect to the beam
    direction.
  }
  \label{fig:ProtonKinematics}
\end{figure}

Figures~\ref{fig:PrimaryLeptonKinematics},~\ref{fig:HadronMultiplicity} and \ref{fig:ProtonKinematics} show distributions from simulated muon neutrino
interactions on carbon nuclei. The neutrinos were generated from muon decay in the production straight and are shown for initial pion energies of 5~GeV and 3~GeV.
Figure~\ref{fig:PrimaryLeptonKinematics} shows the energy of the primary muon and angle of the primary muon to the neutrino beam direction.
The final state visible hadron multiplicity is shown in Figure~\ref{fig:HadronMultiplicity}, and the kinematics of protons produced in these interactions
are shown in Figure~\ref{fig:ProtonKinematics}.

\graphicspath{ {04-Opportunity/Figures/} }

\section{Opportunity}
\label{Sect:Oppo}

nuSTORM will be the first neutrino-beam facility to be based on a
stored muon beam and will provide a test-bed for the development of
the technologies required for a multi-TeV Muon Collider and/or a
Neutrino Factory.
It will also serve the nuclear physics community by providing a unique
probe of flavour-dependent collective effects in nuclei and a new tool
to study the origin of nucleon spin.
Both CERN and FNAL are ideally suited to the implementation of
nuSTORM as the proton infrastructure at each laboratory is well-matched 
to the nuSTORM requirements and the scientific and technology-development
outcomes of nuSTORM are an excellent match to both CERN's and FNAL's
missions.
It is conceivable that the implementation of nuSTORM will drive a
step-change in capability comparable to that produced by Van der
Meer's focusing horn and create a new technique for the study of the
nature of matter and the forces that bind it. 

The ENUBET~\cite{ENUBET:WWW,ENUBET:EUWWW,Torti:2020yzn} and
nuSTORM~\cite{nuSTORM:WWW,Ahdida:2020whw} 
collaborations have begun to work within and alongside the CERN
Physics Beyond Colliders study group~\cite{PBC:WWW} and the
international Muon Collider collaboration~\cite{iMC:WWW} to carry out
a joint, five-year design study and R\&D programme to deliver a
concrete proposal for the implementation of an infrastructure in
which:
\begin{itemize}
  \item ENUBET and nuSTORM deliver the neutrino cross-section
    measurement programme identified in the recent update of the European
    Strategy for Particle Physics and allow sensitive searches for 
    physics beyond the Standard Model to be carried out; and in which
  \item A 6D muon ionisation cooling experiment is delivered as part
    of the technology development programme defined by the
    international Muon Collider collaboration.
\end{itemize}
This document summarises the status of the nuSTORM and
6D-cooling experiments and identifies opportunities for collaboration
in the development of the initiative.

\begin{figure}
  \begin{center}
    \includegraphics[width=0.65\textwidth]{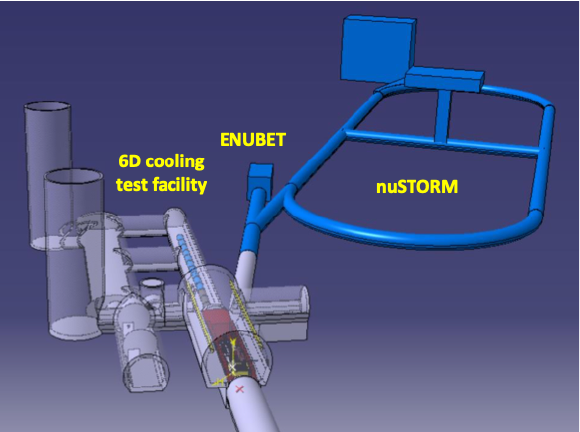}
  \end{center}
  \caption{
    Schematic of a facility capable of serving ENUBET, nuSTORM and a
    6D ionisation cooling facility on the CERN PS.
    The proton beam, entering from the bottom edge of the figure,
    impinges on the meson-production target placed in the neck of a
    horn which focuses accepted beam into a chicane in which the
    desired momentum bite is accepted.
    A magnetic switchyard and transfer lines then transport the
    resulting pion and kaon beams to the three experimental
    facilities~\cite{sauraesteban:ipac2022-thpotk052,Adolphsen:2022bay}.
  }
  \label{Fig:Sct4:Schema}
\end{figure}

Strong synergies have been identified in the proton, target,
meson-capture, and radiation-safety facility required to serve ENUBET
(Enhanced NeUtrino BEams from kaon Tagging;
NP06)~\cite{ENUBET:WWW,ENUBET:EUWWW,Torti:2020yzn},
nuSTORM~\cite{nuSTORM:WWW,Ahdida:2020whw}, and the 6D muon ionisation
cooling experiment.
In the European context the study of a facility capable of serving
ENUBET, nuSTORM and the 6D ionisation cooling demonstration experiment
is mandated in the 2020 Update of the European Strategy for Particle
Physics (ESPP)~\cite{EuropeanStrategyGroup:2020pow}, which recommended
that muon beam R\&D should be considered a high-priority future
initiative and that a programme of experimentation be developed to
determine the neutrino cross-sections required to extract the most
physics from the DUNE and Hyper-K long-baseline experiments.
An initial concept for such a facility in which the target station is
served by the PS proton beam is shown in
figure~\ref{Fig:Sct4:Schema}~\cite{sauraesteban:ipac2022-thpotk052,Adolphsen:2022bay}.

The study of nuSTORM is now being taken forward in the context of the
demonstrator facility required by the international Muon Collider
collaboration that includes the 6D muon ionisation cooling experiment.
The muon-beam development activity is being carried out in close
partnership with the ENUBET collaboration and the Physics Beyond 
Colliders Study Group.
The opportunity, therefore, is to forge an international collaboration
to deliver on a five-year timescale a concrete proposal for the
implementation of an infrastructure in which: 
\begin{itemize}
  \item ENUBET and nuSTORM deliver the neutrino cross-section
    measurement programme identified in the ESPP and allow sensitive
    searches for physics beyond the Standard Model to be carried out;
    and
  \item A 6D muon ionisation cooling experiment is delivered as part
    of the technology development programme defined by the
    international Muon Collider collaboration.
\end{itemize}

\clearpage

\appendix

\section*{\Large Appendix}

\graphicspath{ {10-Appendix/10-01-Simulation/Figures/} }

\section{Simulation}
\label{Appen:Simulation}
The studies are based on preliminary simulations to estimate the
fluxes of neutrinos which might be expected at the front face of a
detector of volume 5\,m x 5\,m x 5\,m placed 50\,m downstream of the end
of the nuSTORM production straight.
This is purpose written code which tracks charged particles from the
production target, down the transfer line  into the production
straight and for multiple turns round the nuSTORM ring.
These particles are assumed to propagate down the optical axis of the
machine and with uniform bending fields and without fringe fields.
The pions decay with the correct distribution of lifetimes and produce
muons and neutrinos with the correct kinematic distributions.
The muons are likewise allowed to decay with the correct lifetime and
the decay products are given the correct kinematic distributions.
The resulting neutrinos are tracked and the position and momentum of
those which cross the plane of the front face of the detector and are
within 10\,m of the detector centre are recorded.
Those within $\pm 2.5\,\text{m}$ in the vertical and horizontal plane
are counted as the flux crossing the front face of the detector. 

In order to improve the accuracy of the simulation, the distributions are
smeared using the results of two other programmes.
The momentum distribution and emittance of the pion beam which leaves
the production region are modelled using FLUKA and these distributions
define the starting conditions of the beam.
The magnetic lattice of the transfer line and production straight are
modelled using BDSIM.
These numbers are used to model the emittance of the beam as it
travels through the transfer line, down the production straight and is
finally captured by the ring. The acceptance of the nuSTORM ring to muons is modelled using the results from 
\cite{Lagrange:2018rt}. This is used to smear the position and momentum of the decay products from both the pion decay and the subsequent muon decay.

\section{Background estimates}
\label{Appen:BkgEstimates}

We have modelled the number of neutrinos we expect from the pion flash in the production straight and the background
from muons which decay before they reach the first bend. We have also modelled the number of neutrinos produced by
muons which are captured in the nuSTORM ring and subsequently decay. There are possible additional backgrounds to these two
sources; we have investigated those and have ignored any background which is less than 1\% of the signal. A summary of
the calculations and reasoning is given below.

\subsection{Decays in the transfer line at $E_{\pi}$=5 GeV}
\begin{figure}
  \centerline{
  \includegraphics[scale=0.45]{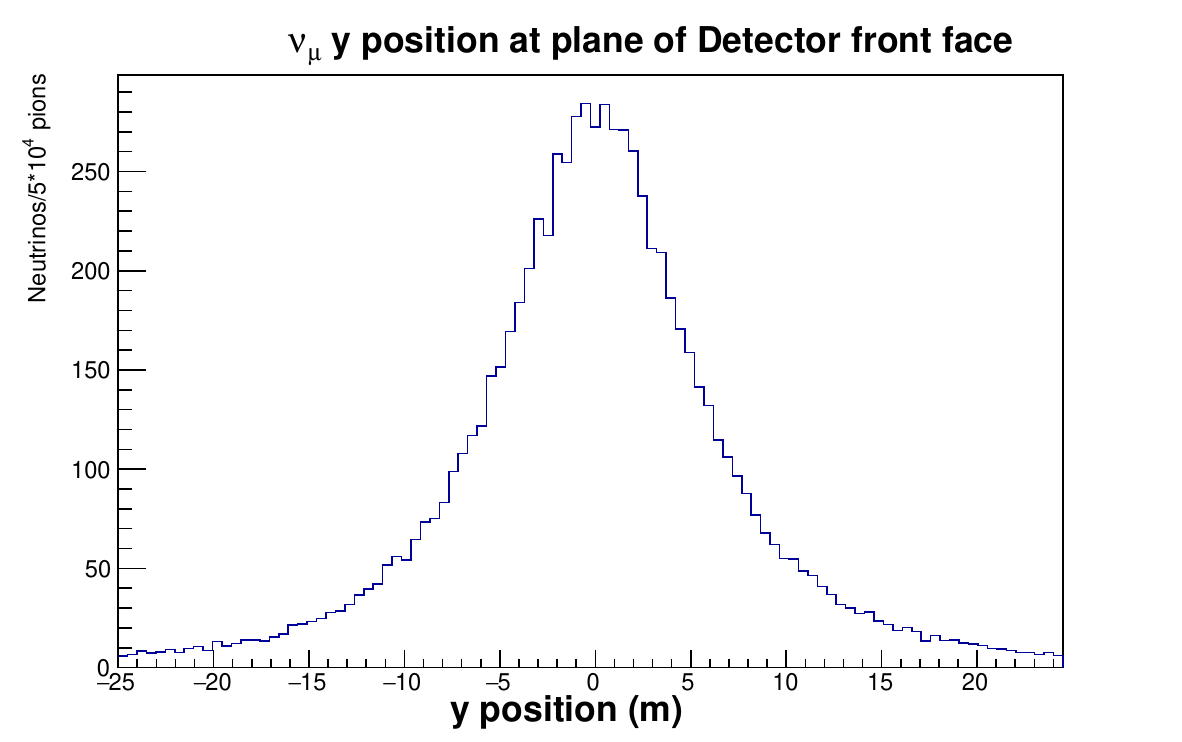}
  \includegraphics[scale=0.45]{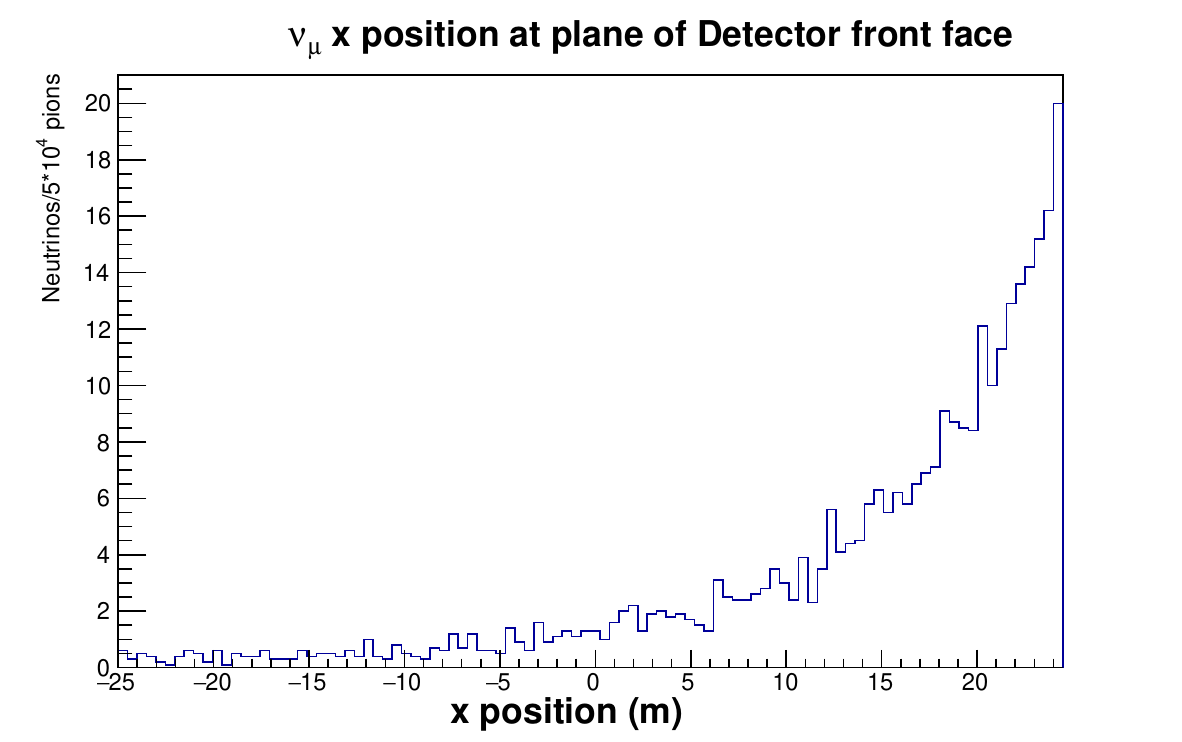}
  }
  \caption{Neutrinos from pion decay in the transfer line at the plane defined by the front face of the detector. Y distribution (left), X distribution (right)}
  \label{fig:E5TLPi}
\end{figure}

Neutrinos from pion decay in the transfer line which reach the detector will overlap in time from the decays in the 
production straight. Figure \ref{fig:E5TLPi} shows the x and y distributions of neutrinos from pion decay in the transfer line at the plane of the detector front face. If we look at the y position (vertical) of the neutrinos which reach the plane, we see that the 
distribution is symmetrical and strongly peaked around zero. The x position has a similar shape, but the transfer line is at an angle
to the production straight and so only the tail of the distribution passes through the detector front  face.  Only 11 neutrinos from 
500k pions at the target arrive at the detector,  compared with the pion flash in the production straight where  7138 neutrinos 
reach the detector from 50k pions; a contamination of less than 0.01\%. The mean energy of these neutrinos is 0.06 GeV; 
in order to  be thrown wide enough by the Q value from the decay to reach the detector, the neutrino must have a low energy 
in the nuSTORM rest frame. This source can therefore be ignored. Of the  81835  muons produced by pion decay 195 decay before the 
end of the transfer line and even if all the neutrinos produced by the muon decay reach  the detector, then this represents 
only 0.27\% of the flash signal. Although some of the muons will make it from the transfer line into the production straight,
most of them will be absorbed by the material surrounding the beam tunnel, or by the material inside the tunnel. They
will be brought to rest by about 15~m of soil. These muons will decay at rest and the resulting neutrinos will be distributed 
isotropically. The acceptance of a detector with a $25~\text{m}^2$ front face at a distance of around 250~m means results in only 8 of these neutrinos being visible in the detector, 
and those at very low energy. We conclude that neutrinos from muon decay in the transfer line can also be ignored.

\subsection{Decays in the transfer line at $E_{\pi}$=3 GeV}

\begin{figure}
  \centerline{
  \includegraphics[scale=0.45]{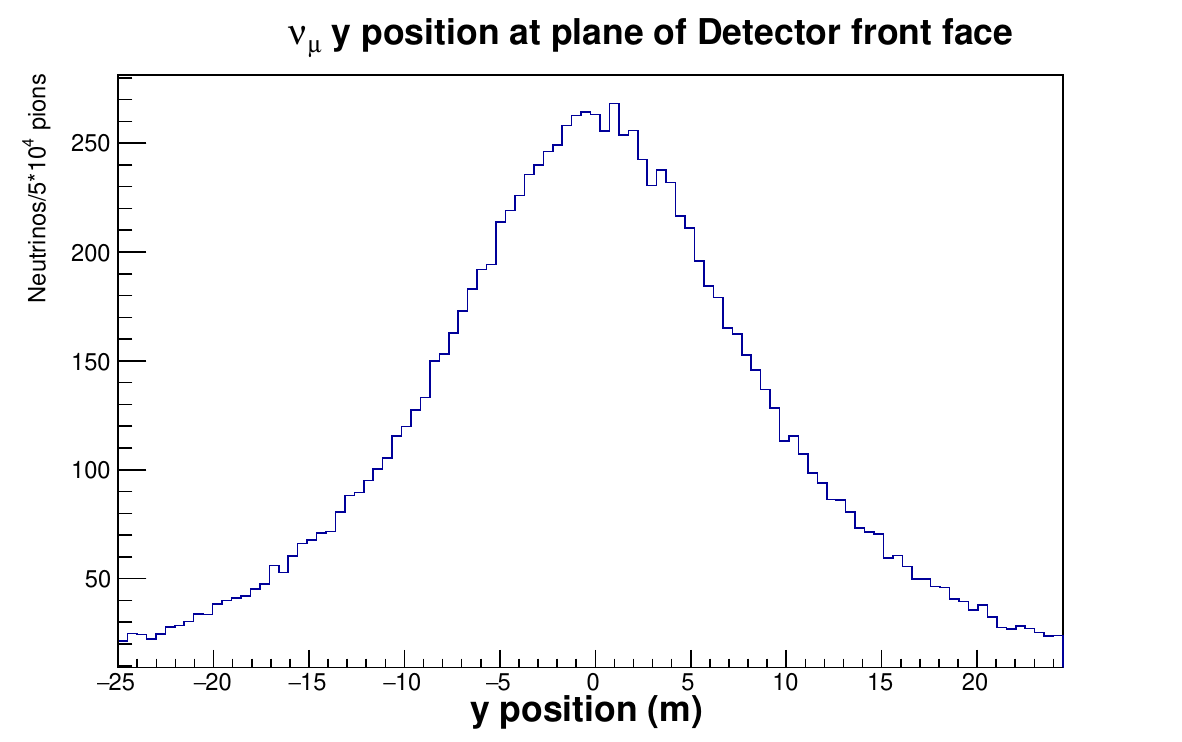}
  \includegraphics[scale=0.45]{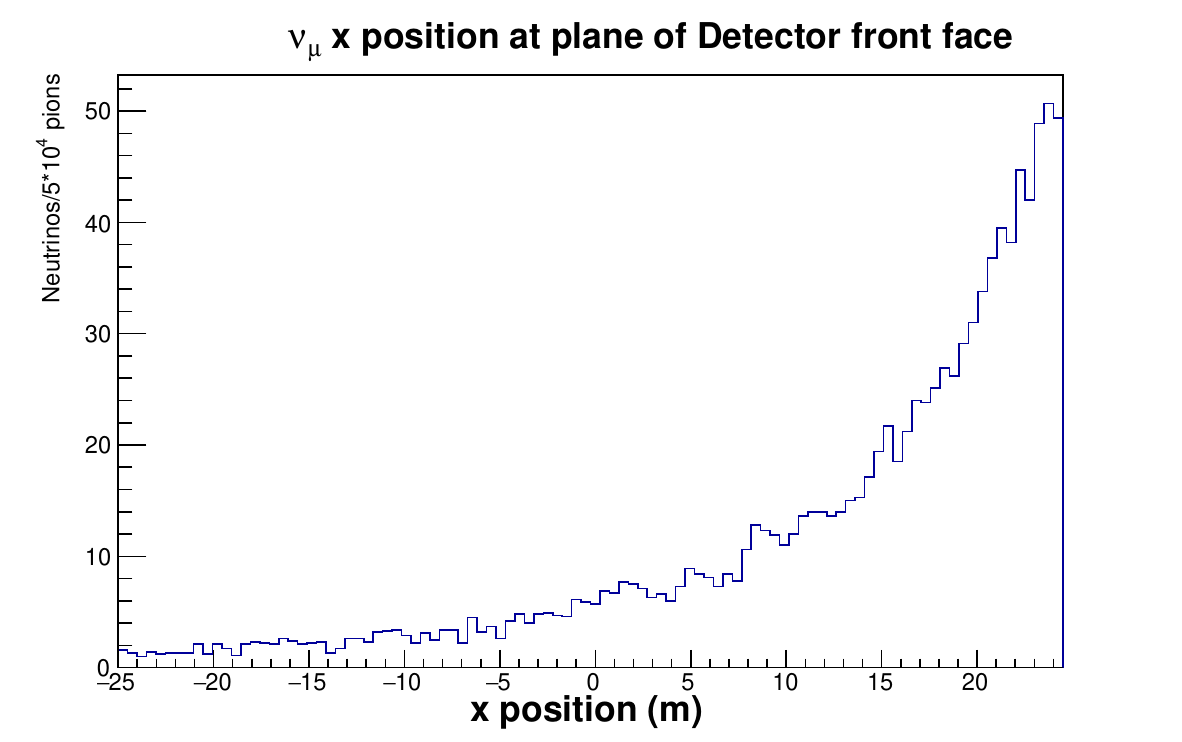}
  }
  \caption{Neutrinos from pion decay in the transfer line at the plane defined by the front face of the detector. Y distribution (left), X distribution (right)}
  \label{fig:E3TLPi}
\end{figure}

At 3 GeV the distribution is wider, as can be seen in Figure \ref{fig:E3TLPi}. The peak at zero is lower in the y position but the tail is higher in the x position.
The combined effect of these is that 34 neutrinos arrive at the detector and their mean energy is slightly higher (0.10 GeV),
because the forward beaming from the pion momentum is smaller and thus the Q value can push the resulting
neutrinos out to wider angles. The contribution at less than 0.03\% is still negligible.
The muon decays in the production straight are still at a low level and even if the transfer line was directed straight at the detector they would still only produce  a rate of around 0.2\%.

The same argument on the muons which are bought to rest by the surrounding material, but in a shorter distance, applies for the 3 GeV sample.

Hence, decays in the transfer line for both pions and muons at pion energies of 3 and 5 GeV are at a level we ignore in this 
study.

\subsection{Decays at the start of the first bend}
We don't have a design which accurately models what happens to the pions as they enter the first bend at the
end of the production straight. However we know from modelling of the transfer line that by the time the
beam has been bent by 8 degrees neutrinos produced by the beam have only a small chance of reaching the
detector. If we look at the number of pions which decay in the the arc as the beam turns through 8 degrees and assume
that the acceptance for these neutrinos is the average of that for a beam at zero degrees and a beam at 8 degrees,
then the background is about 1\%. And since most pions will actually hit an absorber and at worst decay at
rest, the actual background will be smaller and at much lower energies. 

\subsection{Kaons in the transfer line}
We have not modelled Kaon kinematics, but the number produced at the target is significantly lower than the number of pions. 
Even if all of those decay, the number of Kaon decays is less than the number of pion decays and there is no reason to
believe  the acceptance for their neutrinos will be any greater than those from pion decays. We conclude the number
of kaon neutrinos is much less than our 1\% cutoff for modelling.

\newpage
\bibliographystyle{99-Styles/utphys}
\bibliography{Concatenated-bibliography}

\clearpage
\thispagestyle{plain}
\setlength\parindent{0em}

\section*{The nuSTORM collaboration}

\noindent{\bf Canada}
\par \filbreak
  S.~Bhadra, S.~Menary         
  \\{\it
    Department of Physics and Astronomy, York University, 
    4700 Keele Street, Toronto, Ontario, 
    M3J 1P3, Canada
  }\\
  
\par \filbreak
    M.~Hartz$^\dagger$         
  \\{\it
    TRIUMF, 4004 Wesbrook Mall, Vancouver, BC V6T 2A3, Canada
  }
  \\{\small\it
    $^\dagger$ Also at Kavli Institute for the Physics and Mathematics
    of the Universe, The University of Tokyo Kashiwa Campus, 5-1-5
    Kashiwanoha, Kashiwa, Chiba 277-8583, Japan
  } \\

\noindent{\bf China}
\par \filbreak
  J.~Tang         
  \\{\it
    Institute of High Energy Physics, Chinese Academy of Sciences, 
    Beijing, China
  }\\

\noindent{\bf Germany}
\par \filbreak
  U.~Mosel    
  \\{\it
    Justus Liebig Universit\"at, Ludwigstraße 23, 35390 Gießen,
    Germany
  } \\

\par \filbreak
  M.V.~Garzelli         
  \\{\it
    II. Institut f\"ur Theoretische Physik, Universit\"at Hamburg,
    Luruper Chaussee 149, 22761 Hamburg, Germany
  } \\ 

\par \filbreak
  W.~Winter         
  \\{\it
    Deutsches Elektronen-Synchrotron (DESY), Platanenallee 6,
    15738 Zeuthen, Germany
  } \\

\noindent{\bf India}
\par \filbreak
  S.~Goswami, K.~Chakraborty         
  \\{\it
    Physical Research Laboratory, Ahmedabad 380009, India
  } \\ 

\par \filbreak
  S.K.~Agarwalla         
  \\{\it
    Institute of Physics, Sachivalaya Marg, Sainik School Post, 
    Bhubaneswar 751005, Orissa, India
  } \\

\noindent{\bf Italy}
\par \filbreak
  E.~Santopinto         
  \\{\it
  INFN Sezione di Genova, Via Dodecaneso, 33-16146, Genova, Italy
  }\\
  
\par \filbreak
  M.~Bonesini         
  \\{\it
    Sezione INFN Milano Bicocca, Dipartimento di Fisica 
    G.~Occhialini, Milano, Italy
  }\\

\par \filbreak
  L.~Stanco         
  \\{\it
    INFN, Sezione di Padova, 35131 Padova, Italy
  } \\ 

\par \filbreak
  D.~Orestano, L.~Tortora         
  \\{\it
  INFN Sezione di Roma Tre and Dipartimento di Matematica e Fisica, 
  Universit\`{a} Roma Tre, Roma, Italy
  }\\

\newpage
\noindent{\bf Japan}
\par \filbreak    
  Y.~Mori
  \\{\it
    Kyoto University Institute for Integrated Radiation and Nuclear
    Science (KURNS), 2, Asashiro-Nishi, Kumatori-cho, Sennan-gun,
    Osaka 590-0494, Japan
  } \\ 
  
\par \filbreak
  Y.~Kuno, A.~Sato    
  \\{\it
    Osaka University, Graduate School, School of Science, 
    1-1 Machikaneyama-cho, Toyonaka, Osaka 560-0043, Japan
  } \\

\noindent{\bf South Korea}
\par \filbreak
  M.~Chung         
  \\{\it 
    UNIST, Ulsan, Korea
  }\\

\noindent{\bf The Netherlands}
\par \filbreak
  F.~Filthaut$^\dagger$         
  \\{\it
    Nikhef, Amsterdam, The Netherlands
  }\\
  {\small\it
    $^\dagger$ Also at Radboud University, Nijmegen, The Netherlands
  } \\

\noindent{\bf Poland}
\par \filbreak
  J.T.~Sobczyk         
  \\{\it
    Institute of Theoretical Physics, University of Wroclaw, 
    pl. M. Borna 9,50-204, Wroclaw, Poland
  } \\

\noindent{\bf Spain}
\par \filbreak
  J.J.~Gomez-Cadenas         
  \\{\it
    Donostia International Physics Center (DIPC), Paseo Manuel de
    Lardizabal 4, 20018 Donostia-San Sebasti\'an, Gipuzkoa, Spain
  } \\ 
      
\par \filbreak
  J.A.~Hernando~Morata
  \\{\it 
    Universidade de Santiago de Compostela (USC), Departamento de 
    Fisica de Particulas, E-15706 Santiago de Compostela, Spain
  } \\ 

\par \filbreak         
  L.~Alvarez~Ruso, A.~Cervera, A.~Donini, P.~Hernandez, 
  J.~Lopez~Pavon$^\dagger$, J.~Mart\'in-Albo, O.~Mena, P.~Novella,
  M.~Sorel
  \\{\it
    Instituto de Fisica Corpuscular (IFIC), Centro Mixto CSIC-UVEG, 
    Edificio Institutos Investigación, Paterna, Apartado 22085, 46071 
    Valencia, Spain
  }  \\
  {\small\it
    $^\dagger$ Theoretical Physics Department, CERN, 1211 Geneva 23, 
    Switzerland} \\

\noindent{\bf Sweden}
\par \filbreak    
  R.~Ruber
  \\{\it
    Department of Physics and Astronomy, Uppsala University,
    \r{A}ngstr\"omlaboratoriet, L\"agerhyddsv\"agen 1, Box 516,
    751 20 Uppsala, Sweden
  } \\

\newpage
\noindent{\bf Switzerland}
\par \filbreak    
  C.C.~Ahdida, W.~Bartmann, J.~Bauche, M.~Calviani, N.~Charitonidis,
  J.~Gall, B.~Goddard, C.~Hessler, J.~Kopp$^\dagger$, M.~Lamont,
  J.A.~Osborne, E.~Radicioni, A.~de~Roeck, F.M.~Velotti
  \\{\it
    CERN,CH-1211, Geneva 23, Switzerland
  }\\
  {\small\it
    $^\dagger$ Also at PRISMA Cluster of Excellence, Johannes Gutenberg 
    University, Mainz, Germany
  } \\ 
  
\par \filbreak
  A.~Blondel, E.N.~Messomo, F.~Sanchez Nieto    
  \\{\it
    University de Geneve, 24, Quai Ernest-Ansermet, 1211 Geneva 4,
    Suisse
  } \\

\noindent{\bf United Kingdom}
\par \filbreak         
  M.A.~Uchida
  \\{\it
    Cavendish Laboratory (HEP), JJ Thomson Avenue, Cambridge,
    CB3 0HE, UK
  } \\ 

\par \filbreak
  S.~Easo, R.E.~Edgecock, J.B.~Lagrange, W.~Murray, C.~Rogers      
  \\{\it
    STFC Rutherford Appleton Laboratory, Chilton, Didcot, 
    Oxfordshire, OX11 0QX, UK
  } \\ 
  
\par \filbreak         
  J.J.~Back, G.~Barker, S.B.~Boyd, P.F.~Harrison
  \\{\it
    Department of Physics, University of Warwick, Coventry,
    CV4 7AL, UK
  } \\ 
  
\par \filbreak
  S.~Pascoli    
  \\{\it
    Institute for Particle Physics Phenomenology, Department of
    Physics, University of Durham, Science Laboratories, South Rd,
    Durham, DH1 3LE, UK
  } \\ 

\par \filbreak
  S.-P. Hallsj\"o, F.J.P.~Soler    
  \\{\it
    School of Physics and Astronomy, Kelvin Building, 
    University of Glasgow, Glasgow G12 8QQ, Scotland, UK
  } \\ 
  
\par \filbreak
  H.M.~O'Keeffe, L.~Kormos, J.~Nowak, P.~Ratoff    
  \\{\it
    Physics Department, Lancaster University, Lancaster, LA1 4YB, UK
  } \\

\par \filbreak   
  C.~Andreopoulos$^\dagger$, N.~McCauley, C.~Touramanis
  \\{\it
    Department of Physics, Oliver Lodge Laboratory, 
    University of Liverpool, Liverpool, L69 7ZE, UK
  }
  \\{\small\it
    $^\dagger$ Also at STFC, Rutherford Appleton Laboratory,
    Harwell Campus, Chilton, Didcot, OX11 0QX, UK
  } \\ 

\par \filbreak    
  T.~Alves, D.~Colling, P.~Dornan, P.~Dunne, P.~Franchini, P.M.~Jonsson,
  P.B.~Jurj, A.~Kurup, P.~Litchfield, K.~Long$^\dagger$, T.~Nonnenmacher,
  M.~Pfaff$^*$, J.~Pasternak$^\dagger$, M.~Scott, J.K.~Sedgbeer, W.~Shorrock,
  M.O.~Wascko
  \\{\it
    Physics Department, Blackett Laboratory, Imperial College London,
    Exhibition Road, London, SW7 2AZ, UK
  }
  \\{\small\it
    $^\dagger$ Also at STFC, Rutherford Appleton Laboratory,
    Harwell Campus, Chilton, Didcot, OX11 0QX, UK \\
    $^*$ Also at Technical University of Munich (TUM), Arcisstrasse 21
    D-80333 Munich, Germany
  } \\ 

\par \filbreak
  F. di Lodovico, T. Katori    
  \\{\it
    King's College London, Strand, London WC2R 2LS, UK  
  } \\

\par \filbreak
  A.~Bevan, L.~Cremonesi, P.~Hobson    
  \\{\it
    Queen Mary University of London, Mile End Road, London E1 4NS,
    UK
  } \\

\par \filbreak
  R.~Nichol    
  \\{\it
    Department of Physics and Astronomy, University College London, 
    Gower Street, London, WC1E 6BT, UK
  } \\

\par \filbreak
  R.~Appleby, S.~Tygier         
  \\{\it
    The University of Manchester, 7.09, Schuster Laboratory, Manchester, 
    M13 9PL, UK and the Cockcroft Institute, Daresbury Laboratory, 
    WA4 4AD, UK
  } \\

\par \filbreak
  X.~Lu, D.~Wark, A.~Weber$^\dagger$         
  \\{\it
    Particle Physics Department, The Denys Wilkinson Building,
    Keble Road, Oxford, OX1 3RH, UK
  }
  \\{\small\it
    $^\dagger$ Also at STFC, Rutherford Appleton Laboratory,
    Harwell Campus, Chilton, Didcot, OX11 0QX, UK
  } \\ 

\par \filbreak
  P.J.~Smith       
  \\{\it
    University of Sheffield, Dept. of Physics and Astronomy, 
    Hicks Bldg., Sheffield S3 7RH, UK
  } \\

\par \filbreak
  P.~Kyberd, D.R.~Smith   
  \\{\it 
    College of Engineering, Design and Physical Sciences,
    Brunel University London, Uxbridge, Middlesex, UB8 3PH, UK
  } \\

\noindent{\bf United States of America}
\par \filbreak    
  S.J.~Brice, A.D.~Bross, S.~Chattopadhay$^\dagger$, S.~Feher,
  L.~Fields, P.~Hanlet, N.~Mokhov, J.G.~Morf\'in, D.~Neuffer,
  J.~Paley, S.~Parke, Z.~Pavlovic, M.~Popovic, P.~Rubinov, V.~Shiltzev
  \\{\it
    Fermilab, P.O. Box 500, Batavia, IL 60510-5011, USA
  } \\
  {\small\it
    $^\dagger$ Northern Illinois University, 1425 W. Lincoln Hwy.,
    DeKalb, IL 60115-2828, USA} \\ 

\par \filbreak
  P.~Huber, C.~Mariani, J.M.~Link    
  \\{\it
    Virginia Polytechnic Inst. and State Univ., Physics Dept.,
    Blacksburg, VA 24061-0435
  } \\
  
\par \filbreak
  B.~Freemire, A.~Liu         
  \\{\it
    Euclid Techlabs, LLC, 365 Remington Blvd, Bolingbrook, IL, 60440,
    USA
  }\\

\par \filbreak
  D.M.~Kaplan, P.~Snopok    
  \\{\it
  Illinois Institute of Technology, Chicago, IL, USA
  }\\

\par \filbreak
  S.R.~Mishra
  \\{\it
    Department of Physics and Astronomy, 
    University of South Carolina, 
    Columbia SC 29208, USA
  } \\
  
\par \filbreak
  K.~Mahn         
  \\{\it 
    High Energy Physics, Biomedical-Physical Sciences Bldg.,
    Michigan State University, 220 Trowbridge Rd, East Lansing,
    MI 48824, USA
  } \\ 
  
\par \filbreak
  A.~de~Gouv\^ea         
  \\{\it
    Northwestern University, Dept. of Physics and Astronomy, 
    2145 Sheridan Road, Evanston, Illinois 60208-3112 USA
  } \\ 
  
\par \filbreak    
  V.~Pandey
  \\{\it
    Department of Physics, University of Florida, Gainesville, FL
    32611, USA
  }\\

\par \filbreak
  Y.~Onel, D.~Winn         
  \\{\it 
    Department of Physics and Astronomy, The University of Iowa,
    203 Van Allen Hall, Iowa City, Iowa 52242-1479, USA
  } \\
  
\par \filbreak         
  H.A.~Tanaka
  \\{\it
    SLAC National Accelerator Laboratory, 2575 Sand Hill Rd,
    Menlo Park, CA 94025, USA
  } \\ 
  
\par \filbreak
  M.~Hostert         
  \\{\it 
    School of Physics and Astronomy, University of Minnesota,
    Minneapolis, MN 55455, USA 
  } \\

\par \filbreak
  S.A.~Bogacz    
  \\{\it
    Thomas Jefferson National Accelerator Facility, 12000 Jefferson
    Avenue, Newport News, VA 23606, USA
  } \\ 
    
\par \filbreak
  L. Cremaldi, D.~Summers         
  \\{\it
    University of Mississippi, Oxford, MS, USA
  } \\ 

\par \filbreak
  K.T.~McDonald         
  \\{\it
    Princeton University, Princeton, NJ, 08544, USA
  } \\ 

\par \filbreak
  G.~Hanson         
  \\{\it
    Department of Physics and Astronomy, University of California, 
    Riverside, CA 92521, US
  } \\ 

\par \filbreak
  M.~Palmer         
  \\{\it
    Brookhaven National Laboratory, P.O. Box 5000, Upton, NY 11973
    USA
  }\\

\par \filbreak
  M.~Liu         
  \\{\it
    Purdue University, 610 Purdue Mall, West Lafayette, IN, 47907,
    765-494-4600, USA
  }\\

\par \filbreak

\end{document}